\documentclass{mystyle}  

\usepackage{rotating} 
\usepackage{amssymb,amsmath}   
\usepackage{amsfonts} 
\usepackage{bm}

\makeatletter
\def\@dottedtocline#1#2#3#4#5{\ifnum #1>\c@tocdepth \else
  \vskip \z@ \@plus.2\p@
  {\leftskip #2\relax \rightskip \@tocrmarg \parfillskip -\rightskip
    \parindent #2\relax\@afterindenttrue
   \interlinepenalty\@M
   \leavevmode
   \@tempdima #3\relax
   \advance\leftskip \@tempdima \null\hskip -\leftskip
    {#4}\nobreak
        \hfill \nobreak
           \hb@xt@\@pnumwidth{%
             \hfil\normalfont \normalcolor #5}\par}\fi}
\def\numberline#1{\hb@xt@\@tempdima{#1.\hfil}}
\makeatother
\newcommand{\bbox}{\bf}
\begin{document}

\setcounter{chapter}{0}

\chapter{Gluon Radiation and Parton Energy Loss}

\markboth{A. Kovner and U.A. Wiedemann}{Gluon Radiation and Parton Energy Loss}

\author{Alexander Kovner$^{a,c}$ and Urs Achim Wiedemann$^{b,c}$}

\address{$^a$Department of Mathematics and Statistics, University of 
Plymouth,\\ Drake Circus, PL4 8AA, Plymouth, UK\\
$^b$Theory Division, CERN, CH-1211 Geneva 23, Switzerland\\
$^c$Institute for Nuclear Theory, University of Washington, \\ Box 351550,
Seattle, WA 98195, USA}

\begin{abstract}
The propagation of hard partons through spatially extended matter 
leads to medium-modifications of their fragmentation pattern. Here, 
we review the current status of calculations of the corresponding
medium-induced gluon radiation, and how this radiation affects hadronic 
observables at collider energies. 
\end{abstract}
\newpage

\tableofcontents
\newpage
\section{Introduction}
In the QCD improved parton model, inclusive cross sections for
high-$p_\perp$ partons (observed as leading hadrons or jets)
can be calculated through collinear factorization. This
formalism convolutes the perturbatively calculable 
parton-parton cross sections with non-perturbative but
process-independent (``initial state'') parton distribution
functions and (``final state'') fragmentation functions. 
What additional physics input is needed if one extends the
application of QCD to 
the description of inclusive high-$p_\perp$ parton cross 
sections in nucleus-nucleus collisions at RHIC or LHC,
where hadronic projectiles have significant spatial
extension $\propto A^{1/3}$ ? 

The current discussion of this question focuses in 
particular on the following effects:
(i) The density of the incoming parton distributions increases for
given $x$ and $Q^2$ by a factor $\approx A^{1/3}$. This implies that
with increasing $A^{1/3}$ the non-linear modifications
\cite{Gribov:tu,Mueller:wy,Balitsky:1995ub,Jalilian-Marian:1997dw,Kovchegov:1999yj} of the QCD 
evolution equations become relevant at lower center of mass energies 
(or larger momentum fractions $x$). 
(ii) The QCD evolution of the projectile wave function may be
altered due to the presence of a spatially extended ``target'' 
through which it propagates\cite{Bodwin:1988fs}. This goes under the name
``initial state partonic energy loss''. It is discussed e.g.  
in Drell-Yan production in $p-A$ where a quark in the proton
may undergo multiple scattering in the nucleus on its route to
annihilation~\cite{Brodsky:1996nj,Kopeliovich:an}. 
(iii) The hard partonic collision is well-localized on a scale
$1/Q$ which is much smaller than the diameter of a nucleon
$1/\Lambda_{QCD}$. Thus, one does not expect medium-modifications
of the hard partonic cross section itself. However, there is a 
class of high-$p_\perp$ observables in e-A and p-A for which 
the medium-modification can be described by nuclear enhanced
twist-four matrix elements whose importance is determined by
the pole structure of the hard partonic cross section 
\cite{Luo:ui,Luo:np,Qiu:2001hj}. This is the only case of a
nuclear dependence for which higher twist factorization theorems are proven.
iv) At sufficiently large center of mass energy or for 
sufficiently extended projectiles
($A^{1/3} > 1$), the probability of two independent hard parton 
interactions in the same hadronic collision becomes 
significant \cite{Paver:1982yp}. This is e.g. a significant 
background to some of the hadronic Higgs decay channels 
searched for in p-p collisions at the LHC \cite{DelFabbro:2002pw}. 
(v) Finally, the presence of spatially extended matter can
affect the fragmentation and hadronization of hard partons 
produced in nucleus-nucleus 
collisions\cite{Bjorken:1982,Gyulassy:1993hr,Baier:1996kr}. 
This effect is often referred
to as ``final state partonic energy loss'' or ``jet quenching''
and is expected to affect essentially all high-$p_\perp$ hadronic 
observables in heavy ion collisions at collider energies. The 
main motivation for the study of these observables is that the 
degree of nuclear modification may allow for a detailed 
characterization of the hot and dense matter produced in the collision.

All the above mentioned effects should emerge from a theory of QCD 
processes in nuclear matter. The development of such a
theory is still at the very beginning. The high rate at which new
data are becoming available from RHIC, as well as the rapid development 
of new theoretical ideas and phenomenological interpretations  
make it likely that any comprehensive review will be outdated 
very soon. In this situation, we have decided to focus entirely 
on a simple partonic multiple scattering formalism
which underlies at present almost all ``jet quenching'' calculations.
This limited scope excludes many important issues.
In particular, we do not discuss to what extent 
experimental data from RHIC provide indications of partonic
energy loss (see \cite{Gyulassy:2003mc}). 
Moreover, many phenomenologically proposals for
discovering partonic energy loss, such as the measurement of
back-to-back correlations\cite{Bjorken:1982,Wang:1996yh,Wang:2000fq},
the specific behavior of massive $b$-quarks 
\cite{Dokshitzer:2001zm,Lin:1998bd,Lokhtin:2001nh,Djordjevic:2003qk} 
or particle ratios at high $p_\perp$
are not included in the discussion. Finally, we do not
discuss recent alternative formulations of partonic energy
loss, e.g. in the framework of nuclear enhanced higher twist
matrix elements ~\cite{Guo:2000nz,Wang:2002ri}. 

The presentation is organized as follows: in section~\ref{sec2},
we discuss the formalism of eikonal scattering in which 
$S$-matrix amplitudes are determined in terms of eikonal
Wilson lines in the target field. Section~\ref{sec3} discusses
the generalization of this formalism to non-eikonal trajectories
which take the transverse Brownian motion of the partonic
projectile into account. The medium-induced gluon energy
distribution radiated off a hard parton is calculated in
this generalized formalism. Section~\ref{sec4} discusses
the properties of this medium-induced spectrum and section~\ref{sec5}
turns to some applications.

\section{Gluon Bremsstrahlung in the Eikonal Approximation}
\label{sec2}
At high energy, the propagation time of a parton through a 
target is short, partons propagate independently of each other 
and their transverse positions do not change during the propagation. 
The only effect of the propagation is that the wave function of each
parton in the projectile acquires an eikonal phase due to the interaction 
with the target field\cite{Buchmuller:1998jv,Kovner:2001vi}. 
To be specific, we consider a hadronic 
projectile wave function whose relevant degrees of freedom are the 
transverse positions and color states of the partons,
\begin{equation}
  \Psi_{in} = \sum_{\{\alpha_i,{\bf x}_i\}}\, \psi(\{\alpha_i, {\bf x}_i\})\, 
  \vert\{\alpha_i,{\bf x}_i\}\rangle\, .
  \label{2.1}
\end{equation}
The color index $\alpha_i$ can belong to the fundamental,
anti fundamental or adjoint representation of the color $SU(N)$ group, 
corresponding to quark, anti quark or gluon in the wave function. 
We choose the Lorentz frame in which the
projectile moves in the negative $z$ direction. 
Thus at high energy the light cone coordinate $x^+$
of the projectile does not change during the propagation thought the target.
The projectile emerges form the 
interaction region with the wave function
\begin{equation}
  \Psi_{out}={\cal S}\Psi_{in}=
  \sum_{\{\alpha_i,{\bf x}_i\}}\psi(\{\alpha_i, {\bf x}_i\})
  \prod_iW({\bf x}_i)_{\alpha_i \beta_i}\, \vert\{\beta_i,{\bf x}_i\}\rangle\, .
  \label{2.2}
\end{equation}
Here ${\cal S}$ is the $S$-matrix, and the $W$'s are Wilson lines 
along the (straight line) 
trajectories of the propagating particles
\begin{equation}
  W({\bf x}_i)={\cal P}\exp\{i\int dz^-T^aA^+_a({\bf x}_i, z^-)\}
  \label{2.3}
\end{equation}
with $A^+$ - the gauge field in the target and $T^a$ - the generator of 
$SU(N)$ in a representation corresponding to a given parton\footnote{In 
this section we use the light cone gauge $A^-= 0$, 
in which the parton distributions of the {\it projectile} are simply 
expressed in terms of the parton number operators. In this gauge the 
color fields of the target have a large $A^+$ component, 
thus the eikonal $S$ - matrix eq.(\ref{2.3}) 
is given in terms of $A^+$.
The physical observables are of course gauge invariant and can be 
calculated in any gauge. In Appendix A we present the same calculation
in the "target light cone gauge", $A^+=0$, where the only non vanishing 
component of the target field is $A^i$.}. 

The interaction in the target field changes the 
relative phases between components of the wave function and thus 
``decoheres'' the initial state. As a result 
the final state is different from the initial one, and contains 
emitted gluons. 

The component of the outgoing wave function which belongs to the
subspace orthogonal to the incoming state reads
\begin{eqnarray}
  |\delta\Psi\rangle &=& 
  \left[1-|\Psi_{in}\rangle\langle\Psi_{in}|\right]|\Psi_{out}\rangle\, .
  \label{2.4}
\end{eqnarray}
From this, one can calculate for example the inelastic cross section 
which up to the factor of total flux is given by the probability
\begin{equation}
  P_{\rm inel}=|\delta\Psi|^2 = 1-|s|^2\, ,
  \label{2.5}
\end{equation}
where
\begin{equation}
  s=\langle\Psi_{in}|\Psi_{out}\rangle\, .
  \label{2.6}
\end{equation}

The $S$-matrix element in eq.(\ref{2.6}) has to be averaged over the 
distribution of the color fields in the target.
Also, the total and diffractive cross sections can be calculated 
by projecting other components out of $\Psi_{out}$ \cite{Kovner:2001vi}. 

In what follows, we apply the above framework to calculate the
gluon radiation spectrum radiated off a hard quark which propagates
at high energy through a nuclear target. At this point 
we assume the quark to be coming from outside the target rather 
than being produced inside the target in
a hard scattering event. Thus it carries with it the fully developed
wave function which contains the cloud of quasi real gluons. 

In the
first order in perturbation theory the incoming
wave function contains the Fock state $ \vert \alpha\rangle$ 
of the bare quark, supplemented by the coherent state of quasi real 
gluons which build up the Weizs\"acker-Williams field $f({\bf x})$, 
\begin{eqnarray}
  \Psi_{\rm in}^\alpha &=& 
  \vert \alpha\rangle + \int d{\bf x}\,d\xi f({\bf x})\, 
  T^b_{\alpha\, \beta}\, \vert \beta\, ; b({\bf x},\xi)\rangle
  \label{2.7}\\
  && \hspace{-3cm}\epsfxsize=6.0cm 
    \centerline{\epsfbox{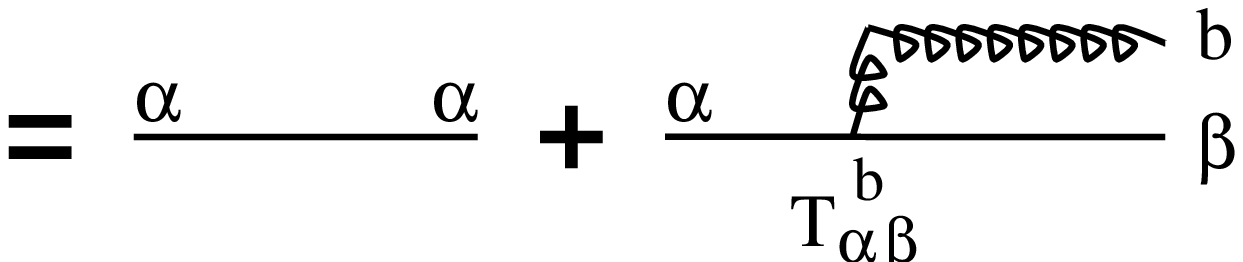}}
  \nonumber
\end{eqnarray}
Here Lorentz and spin indices are suppressed. In the projectile 
light cone gauge $A^-=0$, the gluon field of the projectile is 
the Weizs\"acker-Williams field
\begin{equation}
  A^i({\bf x})\propto  \theta(x^-)\, f_i({\bf x})\, ,\qquad \qquad
  f_i({\bf x})\propto g{{\bf x}_i\over {\bf x}^2}\, ,
  \label{2.8}
\end{equation}
where $x^-=0$ is the light cone coordinate of the quark in the wave 
function. The integration over the rapidity of the gluon in the wave 
function (\ref{2.7}) goes over the gluon rapidities smaller than that 
of the quark. In the leading logarithmic order the wave function 
does not depend on rapidity and
we suppress the rapidity label in the following. 

The interaction of the projectile (\ref{2.7}) with the target leads
to a color rotation $\alpha_i \to \beta_i$ of each projectile
component $i$, resulting in 
an eikonal phase $W({\bf x}_i)_{\alpha_i \beta_i}$. The outgoing
wave function reads
\begin{eqnarray}
 \Psi_{\rm out}^\alpha &=& 
  W^F_{\alpha\, \gamma}({\bf 0})\, \vert\gamma\rangle +
  \int   d{\bf x}\, f({\bf x})\, T_{\alpha\, \beta}^b
  W_{\beta\, \gamma}^F({\bf 0})\, W_{b\, c}^A({\bf x})\, 
  \vert \gamma\, ;c({\bf x})\rangle\, ,
  \label{2.9}
\end{eqnarray}
where  $W^F({\bf 0})$ and $W^A({\bf x})$ are the Wilson lines in the 
fundamental and adjoint representations respectively, corresponding 
to the propagating quark at the transverse position ${\bf x}_q={\bf 0}$ 
and gluon at ${\bf x}_g={\bf x}$. Gluon radiation is an inelastic 
process, for which we have to calculate the projection (\ref{2.4})
\begin{eqnarray}
  &&\vert \delta \Psi_\alpha \rangle =
  \vert \Psi_{\rm out}^\alpha\rangle -
  \sum_\gamma    \vert \Psi_{\rm in}(\gamma)\rangle \langle \Psi_{in}(\gamma) 
   \vert \Psi_{out}^\alpha\rangle
  \label{2.10}\\
&&
\epsfxsize=9.0cm 
\centerline{\epsfbox{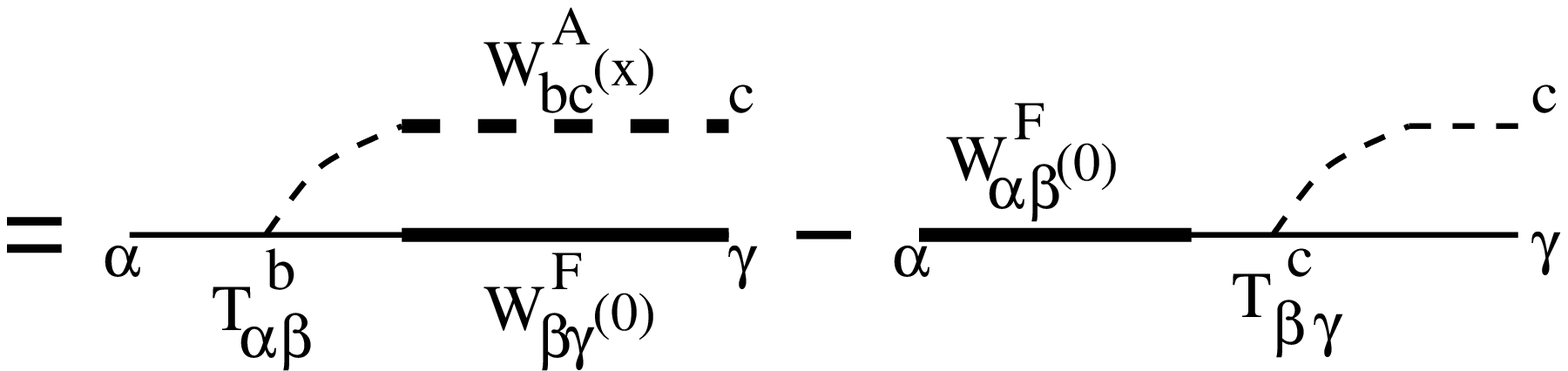}}
\vspace{-1cm}
\nonumber 
\end{eqnarray}
Here, the index $\gamma$ in the projection operator runs over  
the quark color index. It appears in eq.(\ref{2.10}) since
we have to project out the components of the wave function which 
differ from the initial
state only by a global rotation in the color 
space \cite{Kovner:2001vi}\footnote{
An event in which the final state differs from the initial one only 
by a global rotation of the color index 
would be an elastic scattering event.}.
 
The number spectrum of produced gluons is obtained from (\ref{2.10})
by calculating the expectation value of the number operator in the 
state $\delta \Psi_\alpha$, averaged over the incoming color index 
$\alpha$. After some color algebra, one obtains
\begin{eqnarray}
  && N_{\rm prod}({\bf k}) = \frac{1}{N} \sum_\alpha
  \langle \delta \Psi_\alpha\vert a_d^\dagger({\bf k})\, 
  a_d({\bf k})\vert\, \delta\Psi_\alpha \rangle
  \nonumber\\
  && \quad 
  = \frac{\alpha_s\, C_F}{2\pi}\, 
  \int d{\bf x}\, d{\bf y}\, e^{i{\bf k}\cdot({\bf x}-{\bf y})}
  \frac{ {\bf x}\cdot {\bf y}}{ {\bf x}^2\, {\bf y}^2}
  \Bigg[ 1 - \frac{1}{N^2 - 1}\, 
  \langle\langle
  {\rm Tr}\left[ W^{A\, \dagger}({\bf x})\, W^A({\bf 0}) \right]
  \rangle\rangle_t
  \nonumber\\
&& \qquad \qquad \qquad \qquad \qquad \qquad  - \frac{1}{N^2 - 1}\, 
  \langle\langle
  {\rm Tr}\left[ W^{A\, \dagger}({\bf y})\, W^A({\bf 0}) \right]
  \rangle\rangle_t
  \nonumber \\
&& \qquad \qquad \qquad \qquad \qquad \qquad
  +\frac{1}{N^2 - 1}\, 
  \langle\langle
  {\rm Tr}\left[ W^{A\, \dagger}({\bf y})\, W^A({\bf x}) \right]
  \rangle\rangle_t\Bigg]\, .
  \label{2.11}
\end{eqnarray}
Here, we have used $f({\bf x})\, f({\bf y}) = \frac{\alpha_s}{2\pi} \, 
\frac{{\bf x}\cdot {\bf y}}{{\bf x}^2\, {\bf y}^2}$ for the
Weizs\"acker-Williams field of the quark projectile in configuration 
space and the symbol $\langle\rangle_t$ denotes the averaging 
over the gluon fields of the target.

To obtain an explicit expression, we require the target average of the 
product of adjoint Wilson lines. Various averaging procedures are
discussed in the literature
\cite{Zakharov:1996fv,Buchmuller:1998jv,Wiedemann:1999fq,Wiedemann:2000ez,Kovner:2001vi}. 
To be specific, we consider
a target which consists of static scattering centers with scattering
potentials $a_a^+({\bf q})$ at positions $(\hat{\bf x}_n,\hat{z}_n)$. 
In the high energy approximation, each scattering center
transfers only transverse momentum to the projectile,
\begin{equation}
  A_a^+({\bf x},z^-) = \sum_n \int \frac{d^2{\bf q}}{(2\pi)^2}
                       e^{i({\bf x}-\hat{\bf x}_n)\cdot{\bf q}}
                       \, a_a^+({\bf q})\, \delta(z^--\hat{z}^-_n)\, .
\label{2.12}
\end{equation}
We define the target average as an average over the transverse
positions of the static scattering centers. Introducing the
longitudinal density  of scattering centers,
$n(z^-)= \sum_n \delta(z^--\hat{z}^-_n)$, we find
\begin{equation}
    \langle\langle 
    \int dz^-\, d\tilde{z}^-\, A_a^+({\bf x},z^-)\, A_a^+({\bf y},\tilde{z}^-)
  \rangle\rangle_t
  = \int d\xi\, n(\xi)\, \frac{C_A}{2}\, \sigma({\bf x}-{\bf y})\, ,
  \label{2.13}
\end{equation}
where we have defined the dipole cross section 
\begin{equation}
   \sigma({\bf x}-{\bf y}) = 2 \int \frac{d^2{\bf q}}{(2\pi)^2}
   \vert a^+({\bf q})\vert^2\, 
   \left( 1 - e^{i{\bf q}\cdot({\bf x}-{\bf y})}\right)\, .
\label{2.14}
\end{equation}
We now consider the gluon number spectrum (\ref{2.11}) in
two limiting cases.

\subsection{N=1 opacity approximation (single hard scattering limit)}
\label{sec2a}
This limit is relevant for a small target with weak target fields $A^+=O(g)$. 
In this case the projectile can scatter at most once. Thus 
the adjoint Wilson lines can be expanded to leading
order in the single scattering potential $a_a^+({\bf q})$. Using
(\ref{2.13}), we find
\begin{equation} 
  \frac{1}{N^2-1}
  \langle\langle
  {\rm Tr}\left[ W^{A\, \dagger}({\bf x})\, W^A({\bf y}) \right]
  \rangle\rangle_t
  \approx 
  1 - \frac{1}{4\, C_F} \int d\xi\, n(\xi)\, \sigma({\bf x}-{\bf y})\, .
  \label{2.15}
\end{equation}
The corresponding gluon number spectrum reads
\begin{eqnarray}
  N_{\rm prod}({\bf k}) &=& \frac{\alpha_sC_A}{8\pi}\,
  \int d\xi\, n(\xi)\,  
  d{\bf x}\, d{\bf y}\, e^{i{\bf k}\cdot({\bf x}-{\bf y})}
  \frac{ {\bf x}\cdot {\bf y}}{ {\bf x}^2\, {\bf y}^2}
  \Bigg[ \sigma({\bf x}) + \sigma({\bf y}) - \sigma({\bf x}-{\bf y})
  \Bigg]
  \nonumber \\
  &=& \alpha_s\, \pi C_A \int d\xi\, n(\xi)\, 
  \int \frac{d^2{\bf q}}{(2\pi)^2}
   \vert a^+({\bf q})\vert^2\,
   \frac{{\bf q}^2}{{\bf k}^2\, ({\bf k}-{\bf q})^2}\, .
  \label{2.16}
\end{eqnarray}
Differentiated with respect to the momentum transfer ${\bf q}$, 
this result is the well-known Gunion-Bertsch radiation cross 
section\cite{Gunion:qs}
for the gluon radiation cross section in quark-quark
scattering,
\begin{eqnarray}
 && \epsfxsize=9.0cm 
    \centerline{\epsfbox{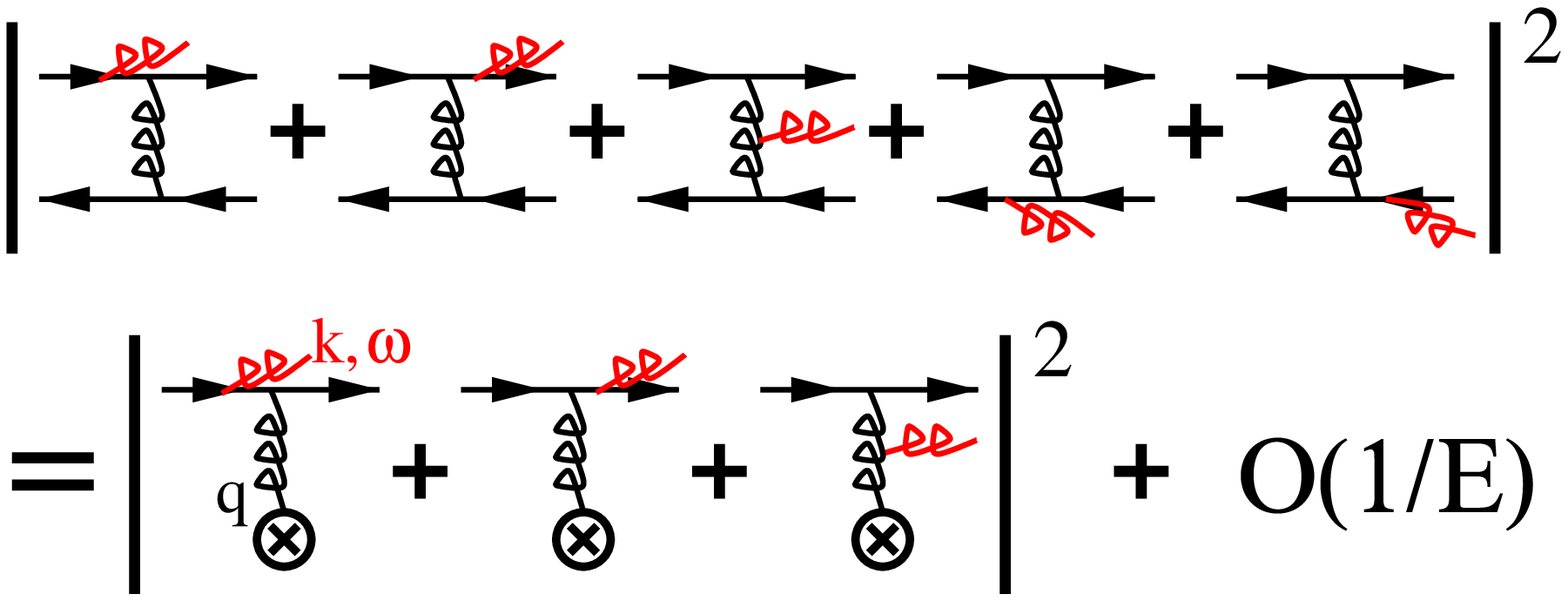}}
\nonumber\\
 &&\qquad \qquad = \frac{d\sigma^{\rm GB}}{d\ln x\, d{\bf q}\, d{\bf k}}
 = \frac{C_A\, \alpha_s}{\pi^2} 
 |a^+_{\rm QCD}({\bf q})|^2\, \frac{{\bf q}^2}{{\bf k}^2\, 
   ({\bf k} - {\bf q})^2}\, .
  \label{2.17}
\end{eqnarray}
Integrated over the transverse momentum exchanged with the
``target'' quark, this radiation spectrum can be written in
terms of the dipole cross section (\ref{2.14}),
\begin{eqnarray}
  \frac{d\sigma^{\rm GB}}{d\ln x\, d{\bf k}}
 &=&  4\, C_A\, \alpha_s \int \frac{d{\bf q}}{(2\, \pi)^2}\,
 \frac{{\bf q}^2}{{\bf k}^2\,
              ({\bf k} - {\bf q})^2}\, 
 \nonumber \\
 && \qquad \qquad \times 
 {-1\over 2}\int {d{\bf r}\over (2\pi)^2}\, 
 \sigma_{\rm QCD}({\bf r})\, 
 \exp\left\{i\, {\bf q}\cdot {\bf r}\right\}\, .
 \label{2.18}
\end{eqnarray}
For a scattering on a static source described by a Yukawa type potential,
one has \footnote{The absolute normalization of 
$a^+$ in eqs.(\ref{2.17}) and (\ref{2.16}) is different, 
which accounts for the different constant factors in these equations.}
$|a^+_{\rm QCD}({\bf q})|^2 \propto
\frac{ \alpha_{\rm s}M^2}{\left(M^2 + {\bf q}^2\right)^2}$. 
Although perturbatively gluons are massless, the 
screening mass $M$ is frequently introduced to mimic the medium 
effects which cut off the infrared divergence\cite{Gyulassy:1993hr}. 
The factor $\alpha_s$ in this expression accounts for the fact that 
the gluon field 
in eqs.(\ref{2.13},\ref{2.14}) is emitted by the single particle 
source. It is the presence of this coupling constant that justifies 
the expansion of the Wilson lines to leading order in $\sigma$.

It is interesting to compare the expression eq.(\ref{2.17}) with 
the high energy limit of the Bethe-Heitler photon radiation cross 
section in QED,
\begin{eqnarray}
 &&\frac{d\sigma^{\rm BH}}{d\ln x\, d{\bf q}\, d{\bf k}}
 = \frac{\alpha_{\rm em}}{\pi^2}\, 
 |a^+_{\rm QED}({\bf q})|^2\, \frac{x^2\, {\bf q}^2}{{\bf k}^2\, 
   ({\bf k} - x{\bf q})^2}\label{2.19} \\
 &&
   \epsfxsize=5.0cm 
    \centerline{\epsfbox{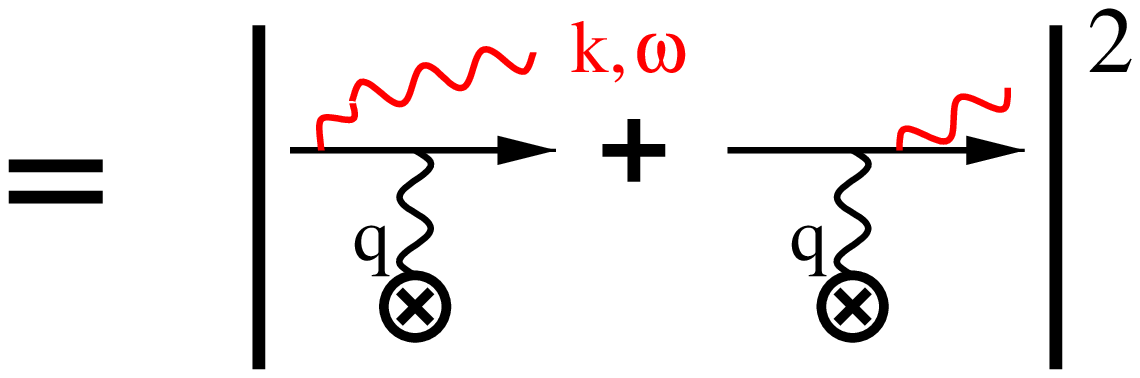}}
    \nonumber
\end{eqnarray}
Here, ${\bf k}$ is the transverse photon momentum and
$x = \omega_\gamma/E$ the fraction of the energy carried away
by the photon. Again, if integrated over the transverse momentum 
transfer ${\bf q}$ from the target, this radiation spectrum
can be written in terms of a dipole cross section 
\begin{eqnarray}
  \frac{d\sigma^{\rm BH}}{d\ln x\, d{\bf k}}
 &&= -2\, \alpha_{\rm em} \int \frac{d{\bf q}\, d{\bf r}}{(2\, \pi)^4}\,
 \frac{{\bf q}^2}{{\bf k}^2\,
              ({\bf k} - {\bf q})^2}\, 
 \sigma_{\rm QED}(x\,{\bf r})\, 
 e^{i\, {\bf q}\cdot {\bf r}}\, .
 \label{2.20}
\end{eqnarray}
The radiation cross section (\ref{2.17}) and (\ref{2.19}) 
for QCD and QED differ only  by their $x$-dependence. The QCD 
radiation is flat in rapidity, whereas the QED radiation cross section
peaks at projectile rapidity $x =1$. The physical reason is that in QCD,
the radiated gluon is charged. The distribution of gluons in the 
incoming quark wave function is flat in rapidity. Thus when a gluon 
scatters directly off a scattering center into the final state, 
the rapidity plateau is produced. It is the 
non-abelian diagram contributing to (\ref{2.17}) which dominates
the QCD radiation spectrum. This is also the reason why the
explicit Casimir factor in (\ref{2.17}) is the adjoint and not
the fundamental one. In QED on the other hand only the electron can 
scatter directly off a source. This scattering is most
effective for electrons with small fraction of longitudinal momentum. 
Such scattering events result in final state photons which carry most 
of the initial longitudinal momentum, and thus the photon
spectrum is peaked around $x=1$. 

In configuration space, the different $x$-dependence of the
QCD and QED radiation spectrum is reflected in the different
arguments of the dipole cross sections: 
the QCD dipole cross 
section is tested at transverse distances ${\bf r}$ but the QED 
one is tested at $x{\bf r}$. For an intuitive understanding
of this behavior, consider the incoming projectile electron or 
quark in the light cone frame as a superposition of the bare 
particle and its  higher Fock states,
\begin{equation}
  \vert {\rm projectile}\rangle = \vert q\rangle + \vert q\, g \rangle
                                 + \dots\, .
  \label{2.21}
\end{equation}
If all Fock components interact with the scattering potentials with the
same amplitude, then the coherence between these amplitudes is not
disturbed, and no bremsstrahlung is generated. The radiation 
amplitude depends on the decoherence between the different Fock state
elements, i.e., on the difference between the elastic scattering 
amplitudes of different fluctuations. This difference 
is characterized by the impact parameter difference between different 
Fock state components. It can be estimated from the transverse  
separation of the $\vert q\rangle$ and $\vert q\,\gamma \rangle$ 
fluctuations. On a characteristic time scale 
${1\over \Delta E} = \left(\frac{k_\perp^2}{2x(1-x)E} \right)^{-1}$ 
set by the transverse energy of the two-particle Fock state, the quark, 
gluon or photon move in the impact parameter plane by distances
\begin{eqnarray}
x_{\perp\, \gamma}\, ,\, x_{\perp\, g} &\propto& {k_\perp\over x\, E}
{1\over \Delta E} \propto {(1-x)\over k_\perp}\, ,
\label{2.22} \\
x_{\perp\, q}\, ,\, x_{\perp\, e} &\propto& {k_\perp\over (1-x)\, E}
{1\over \Delta E} \propto {x\over k_\perp}\, .
\label{2.23}
\end{eqnarray}
Since the photon does not carry charge, the transverse distance
$x_{\perp\, \gamma}$ plays no role in the decoherence of QED
Fock state components, while the distance $x_{\perp\, g}$ plays
the dominant role in the corresponding QCD process.
The QED and QCD dipole cross sections thus measure the transverse 
separations
\begin{eqnarray}
 && \epsfxsize=9.0cm 
    \centerline{\epsfbox{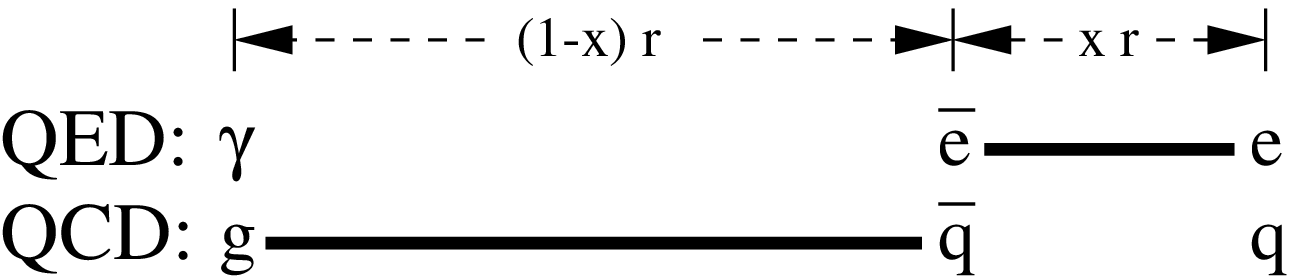}}
\nonumber
\end{eqnarray}
To leading order $O(1/x)$, the cross sections in (\ref{2.18}) and
(\ref{2.20}) thus differ by
a factor $x$ in the argument of $\sigma({\bf r})$. Based on the
above picture, Nikolaev and Zakharov proposed\cite{Nikolaev:1990ja} to 
write the dipole cross section for the $q\bar{q} g$- configuration
beyond leading $O(1/x)$ as 
\begin{eqnarray}
  \sigma_{\rm QCD}({\bf r},x) &=&
  {9\over 8}\, \left\{ \bar{\sigma}({\bf r}) +  
                    \bar{\sigma}((1-x)\, {\bf r})\right\}
  - {1\over 8}\, \bar{\sigma}(x\, {\bf r})\, ,
  \label{2.24}
\end{eqnarray}
where the prefactors account for the color representations
of quark and gluon.

Due to this distinct $x$-dependence
the QED expression cannot be obtained in the eikonal approximation. 
The eikonal limit where the energy loss of the electron is 
neglected is equivalent to the limit $x=0$, where according to eq.(\ref{2.19}) 
no photons are emitted.
In our wave function formalism this arises since the components of the 
wave function with no photons and with one photon rotate with the same phase
during the propagation through the target, if the energies of the electron in 
both components are assumed to be the same. Thus in QED 
the eikonal approximation leads to the vanishing inelastic cross section. 
In QCD this is of course not the case, 
since the gluon component of the wave function scatters most efficiently.

\subsection{Multiple soft scattering limit (dipole approximation)}
\label{sec2b}
We now consider the opposite limit, when 
the projectile undergoes many scatterings. Assume that 
these scatterings are independent in the sense that the vector potentials
of the different scattering centers in eq.(\ref{2.12}) are uncorrelated 
in color space.
Then the dipole cross section in eq.(\ref{2.15}) exponentiates. We also 
assume that the cross section is dominated by the low momentum transfer 
in each individual scattering. In this limit the dipole cross section 
$\sigma({\bf x}-{\bf y})$ can be approximated by the lowest order term 
in the Taylor expansion $\sigma({\bf x}-{\bf y})\propto ({\bf x}-{\bf y})^2$.
This is the so-called logarithmic approximation, in which the
target correlator of two Wilson lines is characterized in terms 
of the saturation momentum $Q_s$,
\begin{eqnarray}
\frac{1}{N^2 - 1}\, 
  \langle\langle
  {\rm Tr}\left[ W^{A\, \dagger}({\bf y})\, W^A({\bf x}) \right]
  \rangle\rangle_t 
  &\approx& \exp\left[- \frac{C_A}{4\, C_F} \int d\xi\, n(\xi)\, 
                     \sigma({\bf x}-{\bf y})\right]
                   \nonumber\\
  &\approx&
  \exp\left[-{({\bf x}-{\bf y})^2\over 8}{C_A\over C_F}Q^2_s\right]\, .  
  \label{2.25}
\end{eqnarray}
Here, the saturation momentum is
\begin{equation}
  Q_s^2 = \int d\xi\, n(\xi)\,   \int \frac{d{\bf q}}{(2\pi)^2}
   \vert a_a^+({\bf q})\vert^2\, {\bf q}^2\, ,
  \label{2.26}
\end{equation}
which in the model (\ref{2.12}) of the target fields characterizes 
the average transverse momentum squared
transferred from the target to the projectile. 
Using the average (\ref{2.25}) to calculate 
the produced gluon number spectrum 
(\ref{2.11}), we obtain~\cite{Kovchegov:1998bi}
\begin{eqnarray}
  N_{\rm prod}({\bf k}) &=& \frac{\alpha_s\, C_F}{2\pi}
        \int d{\bf x}\, d{\bf y}\, e^{ik(x-y)}\,
        \frac{{\bf x}\cdot {\bf y}}{{\bf x}^2\, {\bf y}^2}
        \nonumber \\
        && \times
        \left( 1 + e^{-({\bf x}- {\bf y})^2\, {C_A\over 8C_F}Q_s^2}
        - e^{-{\bf x}^2\, {C_A\over 8C_F}Q_s^2} 
      - e^{- {\bf y}^2\,{C_A\over 8C_F} Q_s^2} \right)
    \nonumber \\
    &=&  \frac{\alpha_s\, C_F}{(2\pi)^3}
    \frac{2\, C_F}{C_A\, Q_s^2} \frac{1}{\pi}\, \int d{\bf q}\, 
    e^{-{2C_F\over C_A Q_s^2}{\bf q}^2}\, 
    \frac{{\bf q}^2}{ {\bf k}^2\, 
    ({\bf k} - {\bf q})^2}\, .
    \label{2.27}
\end{eqnarray}
Expression (\ref{2.27}) contains again the Gunion-Bertsch radiation cross
section (\ref{2.17}). In the dipole approximation (\ref{2.25}),
however, the entire medium acts coherently as a {\it single} 
scattering center with a Gaussian momentum distribution
\begin{equation}
|a_+({\bf q})|^2 \propto e^{-{2C_F\over C_A Q_s^2}{\bf q}^2}\, .
\label{2.28}
\end{equation}
There is an interesting attempt to generalize (\ref{2.27})
to the case of a dense projectile realized in nucleus-nucleus
collisions~\cite{Kovchegov:2000hz}.

The eikonal approximation described in this section is valid as long 
as the average momentum transfer from the medium is much smaller than 
the longitudinal momentum of the produced gluon. Only in this 
limit all scattering centers contribute coherently to the production amplitude.
The purpose of the following is to generalize (\ref{2.27})
to the kinematic region where this is not necessarily the case, 
and incoherent scatterings also play an important role.

\section{Gluon Radiation beyond the Eikonal Approximation}
\label{sec3}
In the eikonal approximation, the  Wilson lines (\ref{2.3}) 
determine the $S$-matrix for the scattering of partonic wave 
functions. In this section, we derive the leading finite energy 
corrections $O(1/p^-)$ to this high energy limit. We explain the
form of the target averages including $O(1/p^-)$ corrections and
we give two applications: the photo absorption cross section and
the medium-induced gluon radiation spectrum. 

\subsection{Wilson Lines for Non-Eikonal Trajectories}
\label{sec3a}
We consider the $N$-fold scattering amplitude of a high energy quark.
In the light cone gauge $A^- = 0$, it reads
\begin{eqnarray}
F(x_0,p) &=& \nonumber \\
&&
\epsfxsize=9.0cm 
\hspace{-2cm}\centerline{\epsfbox{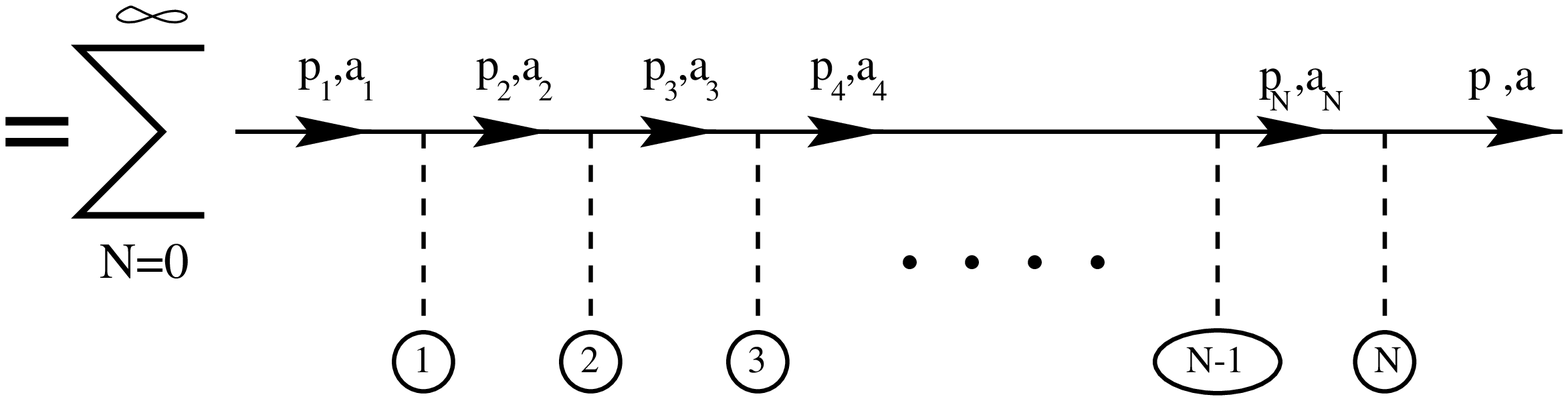}}
\vspace{-1cm}
\nonumber \\
    &&=\sum_{N=0}^{\infty} \prod_{i=1}^N {\cal P}\, \left( \int 
    \frac{dp_i^+\, dp_i^-\, d{\bf p}_i}{(2\pi)^4}
    \int dx_i^+\, dx_i^-\, d{\bf x}_i\, 
    \frac{i}{\not p_i - m + i\epsilon}\, 
    \right.
    \nonumber \\
    && \times 
    \left. 
    \left( -i \gamma^-\, A^+({\bf x}_i,x_i^-)\right)\, 
    e^{i x_i^+\cdot(p^-_{i+1}-p^-_i) +i x^-_i\cdot(p^+_{i+1}-p^+_i)} 
    e^{i {\bf x}_i\cdot({\bf p}_{i+1}-{\bf p}_i)} 
    \right)\, 
    \nonumber \\
    && \times  e^{ix_0\cdot p_1}\, v(p)\, .
  \label{3.1}
\end{eqnarray}
Here, the target gauge fields $A^+ = A^+_a T_a$ are path-ordered 
along the trajectory of the quark; we suppress color and spin 
indices on the quark propagators and the spinor $v(p)$. The 
additional phase factor $e^{ix_0\cdot p_1}$ in (\ref{3.1}) is useful in order
to connect the multiple scattering diagram to the vertex of a more 
complicated amplitude by integration over $x_0$. In the absence 
of scattering centers, (\ref{3.1}) reduces to a free quark
wave function $e^{ix_0\cdot p}\, v(p)$.

We approximate eq. (\ref{3.1}) to leading order 
$O\left( \left(1/p^-\right)^0\right)$ in the norm and to 
subleading order $O\left(1/p^-\right)$ in the phase. To this order 
we can neglect the dependence of the target fields on  $x_i^+$, as 
indicated in (\ref{3.1}). This is because the incoming quark propagates 
in the negative $z$ direction and thus only feels the fields at $x^+=0$.
The integration over $x_i^+$ and $p_i^-$ is then trivial: all propagators
have the same component $p_i^- = p^-$. To simplify (\ref{3.1}) further,
we use 
\begin{equation}
  \frac{1}{\not p - m} = \frac{ \sum_s u_s(p)\, \bar{u}_s(p)}{p^2 - m^2}
                         + \frac{\gamma^-}{2p^-}\, ,
  \label{3.2}
\end{equation}
and neglect the last term to leading order in the energy $1/p^-$. 
The spinor structure can be simplified:
\begin{equation}
 \bar{u}_s(p) \left( -i \gamma\cdot A \right) u_{s'}(p) \simeq p^-\, A^+\, 
 \delta_{ss'}\, .
 \label{3.3}
\end{equation}
With these simplifications, eq. (\ref{3.1}) takes the form
\begin{eqnarray}
    F(x_0,p) &=&   
    \sum_{N=0}^{\infty} \prod_{i=1}^N {\cal P}\, \left( \int 
    \frac{dp_i^+\, d{\bf p}_i}{(2\pi)^3}
    \int dx_i^-\, d{\bf x}_i\, 
    \frac{-i}{ p^+_i - ({\bf p}_i^2+ m^2)/p^- + i\epsilon}\, 
    \right.
    \nonumber \\
    && \times 
    \left. 
    \left( i A^+({\bf x}_i,x_i^-)\right)\, 
    e^{i p_i^+\cdot(x^-_{i-1}-x^-_i)} 
    e^{-i {\bf p}_i\cdot({\bf x}_{i-1}-{\bf x}_i)} 
    \right)\, 
    \nonumber \\
    && \times  e^{ix^+_0\cdot p^-} 
    e^{ix_N^- p^+ - i{\bf x}_N\cdot {\bf p}}\, v(p)\, .
  \label{3.4}
\end{eqnarray}
The spatial ordering of the scattering centers along the
trajectory is fixed explicitly by performing the $p^+_i$-integrations.
Neglecting the mass term, we find a product of $\Theta$-functions
times phase factors
\begin{eqnarray}
  &&\int \frac{dp_i^+}{2\pi}\, 
    \frac{i}{ p^+_i - ({\bf p}_i^2+ m^2)/p^- + i\epsilon}\, 
    e^{i p_i^+(x^-_{i-1}-x^-_i)}
    \nonumber \\
    && \quad =  e^{i \frac{{{\bf p}_i}^2}{p_i^-}\, (x^-_{i-1}-x^-_i)}\, 
    \Theta(x^-_i-x^-_{i-1})\, .
    \label{3.5}
\end{eqnarray}
Inserting this expression into (\ref{3.4}) and performing the 
transverse ${\bf p}_i$-integration, we obtain the basic building block
of the formalism, the free light-cone Green's function
\begin{eqnarray}
 &&G_0({\bf x}_{i-1},x^-_{i-1};{\bf x}_i,x^-_i\vert p^-)
  = \int \frac{d{\bf p}_i}{(2\pi)^2}
  e^{i \frac{{\bf p}_i^2}{p_i^-}\, (x^-_{i-1}-x^-_i)}\, 
  e^{-i {\bf p}_i\cdot({\bf x}_{i-1}-{\bf x}_i)} 
  \nonumber \\
  && = \frac{p^-}{4\pi i(x^-_i-x^-_{i-1})}\, 
  \exp\left[
  {\frac{ip^- ({\bf x}_i-{\bf x}_{i-1})^2}{4\, (x^-_i-x^-_{i-1})}}
  \right]
  \nonumber\\
  && = \int {\cal D}{\bf r}(\xi)\, 
       \exp\left[ \frac{ip^-}{4} \int_{x^-_{i-1}}^{x^-_i}
                  d\xi\, \dot{\bf r}^2(\xi)\right]\, .
  \label{3.6}
\end{eqnarray}
Here, $\dot{\bf r} = d{\bf r}/dz^-$ and the boundary conditions on the
free path integral are ${\bf r}(x^-_i) = {\bf x}_i$. The limit
$x_i^- - x_{i-1}^- \to 0$ of (\ref{3.6}) is equivalent to the high-
energy limit 
\begin{equation}
  \lim_{p^-\to \infty} 
      G_0({\bf x}_{i-1},x^-_{i-1};{\bf x}_i,x^-_i\vert p^-)
  = \delta^2({\bf x}_{i-1}-{\bf x}_i)\, .
  \label{3.7}
\end{equation}
The free Green's function (\ref{3.6}) evolves the free plane wave
$e^{ix_i^- p^+ - i{\bf x}_i\cdot {\bf p}}$ to the
position $({\bf x}_{i-1},x^-_{i-1})$. The sum of the $N$-fold
scattering amplitudes (\ref{3.1}) reads
\begin{eqnarray}
    F(x_0,p) &=& e^{ix^+_0\cdot p^-}    
    \sum_{N=0}^{\infty} {\cal P}\, 
    \Bigg( \prod_{i=1}^N 
    \int dx_i^-\, d{\bf x}_i\, 
    G_0({\bf x}_{i-1},x^-_{i-1};{\bf x}_i,x^-_i\vert p^-)
    \nonumber \\
    && \qquad \times \Theta(x_i^- - x_{i-1}^-)\, 
    \left( i A^+({\bf x}_i,x_i^-)\right)\, \Bigg)\, 
    e^{ix_N^- p^+ - i{\bf x}_N\cdot {\bf p}}\, v(p)
    \nonumber \\
    &=& e^{ix^+_0\cdot p^-} \hspace{-0.2cm} \int d{\bf x}
        G({\bf x}_0,x_0^-;{\bf x},x^-\vert p^-)\, 
          e^{i \frac{{{\bf p}_i}^2}{p_i^-}\, x^-\, 
    -i {\bf p}_i\cdot {\bf x}} \hspace{-0.1cm} v(p)\, ,
  \label{3.8}
\end{eqnarray}
where the full Green's function is
\begin{eqnarray}
  && G({\bf x}_0,x_0^-;{\bf x},x^-\vert p^-)
  \nonumber \\
  &&\quad =  \int {\cal D}{\bf r}(\xi)\, 
       \exp\left[ \frac{ip^-}{4} \int_{x^-_{i-1}}^{x^-_i}
                  d\xi\, \dot{\bf r}^2(\xi)\right]\, 
      W([{\bf r}];x_0^-,x^-)\, ,
  \label{3.9}\\
   && W([{\bf r}];x_0^-,x^-) = {\cal P}\, 
   \exp\left[ i\int_{x_0^-}^{x^-} d\xi\, A^+({\bf r}(\xi),\xi)\right]\, .
  \label{3.10} 
\end{eqnarray}
These expressions give the leading $O(1/p^-)$ corrections to the
phase of the eikonal Wilson line (\ref{2.3}). 
The path integral in (\ref{3.9}) has a very simple interpretation.
It describes the quantum mechanical particle which, while propagating 
through the medium can move in the transverse plane. The free transverse 
motion is governed by the  kinetic term
$\frac{p^-}{4} \dot{\bf r}^2$ - a two-dimensional light-cone hamiltonian
with ``mass'' given by the total energy $p^-/2$ of the projectile 
along the beam. During the motion the wave function of the particle also 
acquires a phase given by the Wilson line along the trajectory. The infinite
energy limit corresponds to the classical limit, where a single straight 
classical trajectory dominates the sum over paths. In this limit the 
Green's function (\ref{3.9}) reduces to the eikonal Wilson line in (\ref{2.3}),
\begin{eqnarray}
  \lim_{p^-\to \infty} 
  G({\bf x}_0,x_0^-;{\bf r},x^-\vert p^-) =   
      W({\bf r};x_0^-,x^-)\, \delta^2({\bf x}_0 - {\bf r})\, .
  \label{3.11} 
\end{eqnarray}
%

\subsubsection{Non-abelian Furry approximation (target field $A_0$)}
\label{sec3a1}

The multiple scattering diagram (\ref{3.8}) contains the leading
order $O(1/p^-)$ corrections to the phase of the eikonal Wilson
line, but the norm of (\ref{3.8}) is given to $O\left((1/p^-)^0\right)$
only. For a scalar quark that would be the only important correction. 
For a spinor quark however this is not enough if one wants for example 
to discuss emission of on shell gluons. It turns out that the leading order
of this propagator vanishes when connected to an emission vertex of a 
real gluon. To get a leading order result for this process one has to 
keep order $O(1/p^-)$ corrections to the first propagator 
$1/(\gamma\cdot p_1 - m)$.
We now discuss the non Abelian 
Furry approximation which addresses this issue.

We also change our notations slightly in accordance with literature. 
The target field will not be described by a large
$A^+$-component but rather specified by the Gyulassy-Wang 
model~\cite{Gyulassy:1993hr,Wang:1994fx} as a collection of scattering 
potentials of Yukawa-type with Debye screening mass $M$,
\begin{eqnarray}
  A_\mu({\bf x}) &=&
  \delta_{0\mu}\, \sum_{i=1}^\infty\, \varphi^a_i({\bf x})\, T^a\, ,
  \label{3.12} \\
  \varphi^a_i({\bf x}) &=& \varphi({\bf x}-\check{\bf x}_i)\, 
  \delta^{a\, a_i}\, .
  \label{3.13}
\end{eqnarray}
Here, the $i$-th scattering center is located at $\check{\bf x}_i$ 
and exchanges a specific colour charge \footnote{Intuitively one thinks 
of large $A^+$ as being a more appropriate description of a target which 
itself is moving in the positive $x_3$ direction, while large $A_0$ being 
more appropriate for a static target. Technically, however there is no 
difference between the two sets of vector potentials as long as the 
projectile is very energetic and its $x^+$ coordinate does not change
during the propagation inside the target. In this sense switching to 
large $A_0$ can be considered as a different gauge fixing.} $a=a_i$. 

For this target field (\ref{3.12}), the calculation of the multiple 
scattering diagram (\ref{3.1}) proceeds in close analogy to section 
\ref{sec3a}. The components $\gamma^-A^+$ in (\ref{3.1}) are replaced 
by $\gamma_0A_0$. The role of the $p^+_i$-integrals (\ref{3.5}) is 
played by the $p_l$-integrals which lead to an ordering along the
longitudinal direction. To simplify the spinor structure, one
uses $(\not p)\gamma_0 \simeq E(\gamma_0 - \gamma_3)\gamma_0$ 
which can be iterated, $\left(E(\gamma_0-\gamma_3)\gamma_0\right)^n
v(p) = 2^{n-1}\, E^n\, (\gamma_0-\gamma_3)\gamma_0\, v(p)
\simeq 2^n \, E^n\, v(p)$. 

In comparison to section \ref{sec3a}, the only new feature of
the present discussion is that we approximate the first quark propagator 
to $O(1/E)$. In a coordinate system with ${\bf p} \parallel {\bf n}$
and for  $\vec{\alpha} = \gamma_0\, \vec{\gamma}$, we have
\begin{eqnarray}
  \not{p}_i\gamma_0 &=& p(\gamma_0-\gamma_3)\gamma_0
  + \alpha^\perp\cdot({\bf p}_1 - {\bf p})
   - \alpha^L\, \left( {{\bf p}_1^2\over 2\, E} -
                   {{\bf p}^2\over 2\, E} \right) 
  \nonumber \\
  \longrightarrow && 
  p\, (\gamma_0-\gamma_3)\gamma_0 + i\vec{\alpha}\cdot
  {\partial\over \partial {\bf y}}
  - {\vec{\alpha}}\cdot\left( {\bf p} - p\, {\bf n}\right)\, .
  \label{3.14}
\end{eqnarray}
Here, the operator ${\partial\over \partial {\bf y}}$ acts 
on wave functions of the form 
$\exp\left\{- i\,{\bf p}_i\cdot{\bf y}\right\}$.
This structure leads to the differential operator $\hat D_i$
 \begin{eqnarray}
 \hat D_i &=&
  1- i\,\frac{\vec{\alpha}\cdot\vec{\nabla}}{2\,E} 
   - \frac{\vec{\alpha}\cdot({\bf p}_i-{\bf n}\,p_i)}{2\,E}\, ,
   \label{3.15}
\end{eqnarray}
which carries the transverse and longitudinal spinor structure to 
order $O(1/E)$. To lowest non-trivial order $O(1/E)$ in the norm 
and phase, the multiple scattering diagram (\ref{3.1}) takes now 
the form\cite{Wiedemann:2000ez}
\begin{eqnarray}
  {\Psi_v}(y^0,{\bf y},p) 
  &\equiv& e^{i\, E\, y^0 - i\, p\, y_L}\, {\hat D}\,
  F({\bf y},{\bf p})\, u({\bf p})\, ,
  \label{3.16}\\
&&
\epsfxsize=9.0cm 
\hspace{-2cm}\centerline{\epsfbox{quarkrescatt.eps}}
\vspace{-1cm}
\nonumber 
\end{eqnarray}
Here, the differential operator ${\hat D}$ acts on the transverse 
wavefunctions $F$ which for far forward longitudinal distances 
$x_L$ satisfies plane wave boundary conditions
\begin{eqnarray}
  F_\infty({\bf x},x_L,{\bf p}_i) &=& 
  \exp\left\{- i\,{\bf p}_i\cdot{\bf x}
  + i\, { {\bf p}_i^2\over 2\, p_i} x_L\right\}\, ,
  \label{3.17} \\
  F({\bf y},y_L,{\bf p}_i) &=& 
  \int d{\bf x}\, 
 \bar G({\bf y},y_L;{\bf x},x_L\vert p) \, 
  F_\infty({\bf x},x_L, {\bf p})\, .
  \label{3.18}
\end{eqnarray}
The full Green's function in this expression is given in 
close analogy to (\ref{3.9}) by the sum over paths weighted 
with the Wilson lines. It satisfies the recursion 
relation involving the free non-interacting Green's function $G_0$,
\begin{eqnarray}
  \bar G({\bf r},z;{\bf r}',z'|p) &\equiv&  G_0({\bf r},z;{\bf r}',z'|p)
  \nonumber \\
  && -i\int\limits_{z}^{z'} d\xi\, 
  \int d{\bbox \rho}\, G_0({\bf r},z;{\bbox \rho},\xi|p)\, 
  A_0({\bbox \rho},\xi)\, 
  G_0({\bbox \rho},\xi;{\bf r}',z'|p)\,
  \nonumber \\
  && + {\cal P} \int\limits_{z_L}^{x_L} d\xi_1\, 
       \int\limits_{\xi_1}^{x_L} d\xi_2\, 
       \int d{\bbox \rho}_1\, d{\bbox \rho}_2\,
       G_0({\bf r},z;{\bbox \rho}_1,\xi_1|p)\, i\, 
  A_0({\bbox \rho}_1,\xi_1) 
  \nonumber \\
  && \quad \times   
  G_0({\bbox \rho}_1,\xi_1;{\bbox \rho}_2,\xi_2|p)\, 
  i\, 
  A_0({\bbox \rho}_2,\xi_2)\, 
  \bar{G}({\bbox \rho}_2,\xi_2;{\bf r}',z'|p)\, ,
  \label{3.19}\\
  G_0({\bf r},z;{\bf r}',z'|p) &\equiv& \frac{p}{2\pi i(z'-z)}
  \exp\left\{ \frac{ip\, \left({\bf r}-{\bf r}'\right)^2}{2(z'-z)} 
              \right\}\, .
  \label{3.20}
\end{eqnarray}
The path ordering ${\cal P}$ in (\ref{3.19}) ensures that the 
potentials $A_0$ are ordered along the longitudinal axis.
%

%
This formula is in fact a direct non Abelian generalization 
of the 
corresponding expression in Quantum Electrodynamics.
Consider the solution of the Dirac equation 
for an electron in a spatially extended scattering potential
$U({\bf x})$ 
\begin{eqnarray}
  && \left[i\, {\partial\over \partial t} - U({\bf x}) - m\,\gamma_0 +
  i\,\bbox{\alpha}\cdot\bbox{\nabla}\right]\,
  \Psi(x,p) = 0\, .
  \label{3.21}
 \end{eqnarray}
In the high-energy approximation in which 
we neglect corrections of order  $O(U/E_i)$ and $O(1/E_i^2)$,
the solution to this Dirac equation is the Furry wave 
function~\cite{Kopeliovich:1998nw,Wiedemann:1999fq},
\begin{eqnarray}
  &&\Psi_F(x,{\bf p}_i) = e^{iE_it - i\vert {\bf p}_i\vert z}\,
   \hat{D}_i\,F({\bf x},{\bf p}_i)\,
   v({\bf p}_i)\, .
   \label{3.22}
 \end{eqnarray}
This solution has the same form as the non-abelian
multiple scattering diagram in (\ref{3.16}). The Green's function
used to evolve the wave function $F$ is now given in terms of the 
scattering potential $U({\bf x})$,
\begin{eqnarray}
  &&\qquad\qquad G({\bf r}, z; {\bf r}',z'|\,p) =  \nonumber \\
  &&\int {\cal D}{\bf r}(\xi)\,
  \exp\left\{ \int\limits_{z}^{z'}{\it d}\xi\,
  \left[\frac{ip}{2} \dot{\bf r}^2(\xi) -
  i\,U\bigl({\bf r}(\xi),\xi\bigr)\right] \right\}\, .
  \label{3.23}
\end{eqnarray}
Again, it describes the quantum mechanical evolution of the projectile 
particle in the plane transverse to the beam. 

\subsection{Target averages for non-eikonal Wilson lines}
\label{sec3b}
In section~\ref{sec2}, we discussed the target average of two
eikonal Wilson lines. In section~\ref{sec3a} we showed that the
$O(1/E)$-improved generalization of eikonal Wilson lines are given
in terms of the path integral (\ref{3.9}). We can compute the
target average of two such Green's functions
expanding both Green's functions in powers of
$A_0$ according to (\ref{3.19}) and then resumming the series. We assume 
that the different scattering centers in (\ref{3.12}) are uncorrelated 
and that the projectile can interact with a given scattering center 
only once (elastically or inelastically). For generality we take the 
two Greens' function to act on initial states with different
energies $(1-\alpha)p$ and $\alpha p$, respectively. This is a generic 
situation in the application to deeply inelastic scattering, where the 
initial quark and anti quark share the energy in the initial state.
This will also be useful for us in the following, since we will need 
to calculate phase differences between the wave functions of an initial 
state parton and a final state parton with a different energy.  
Performing the target average as in (\ref{2.13}), the target average 
for two adjoint Green's functions\cite{Wiedemann:2000za} leads to the 
appearance of the dipole cross section \footnote{In the rest 
of this article we use $\sigma$ to denote the cross section for scattering 
of dipole consisting of two adjoint charges. It differs from $\sigma$ 
defined in (\ref{2.13}) by the factor ${C_A\over C_F}$.} 
$\sigma$,
\begin{eqnarray}
  &&\langle\langle   \bar G({\bf r}_1,z;{\bf r}_1',z'(1-\alpha) p)|
    \bar G({\bf r}_2,z;{\bf r}_2',z'|\alpha p) \rangle\rangle_t
  \nonumber \\
    && \qquad = 
  \int\limits_{{\bf r}_1(z)={\bf r}_1}^{{\bf r}_1(z')={\bf r_1'}} 
  {\cal D}{\bf r}_1\, 
  \int\limits_{{\bf r}_2(z)={\bf r}_2}^{{\bf r}_2(z')={\bf r_2'}}
  {\cal D}{\bf r}_2
     \exp\left[ i\int_{z}^{z'} d\bar{\xi}\, \frac{p}{2}
     \left( \alpha\, \dot {\bf r}_2^2 + (1-\alpha)\, \dot {\bf r}_1^2
       \right) \right.
  \nonumber \\
  && \qquad \qquad \qquad \qquad \left. + i\, \frac{1}{2}\, n(\xi)\, 
       \sigma\left({\bf r}_1(\xi) - {\bf r}_2(\xi)\right)
       \right] \, .
  \label{3.24}
\end{eqnarray}
With the help of the transformation
\begin{eqnarray}
  {\bf r}_a(\xi) &=& (1-\alpha)\, {\bf r}_1(\xi) + \alpha\, {\bf r}_2(\xi)\, ,
  \label{3.25}\\
  {\bf r}_b(\xi) &=& {\bf r}_1(\xi) - {\bf r}_2(\xi)\, ,
  \label{3.26}
\end{eqnarray}
the target average can be written as  
\begin{eqnarray}
  &&\langle\langle   \bar G({\bf r}_1,z;{\bf r}_1',z'(1-\alpha) p)|
    \bar G({\bf r}_2,z;{\bf r}_2',z'|\alpha p) \rangle\rangle_t
  \nonumber \\
  && = {\cal K}_0\bigl({\bf r}_a(z'),z';{\bf r}_a(z),z|p\bigr)\, 
       {\cal K}\bigl({\bf r}_b(z'),z';{\bf r}_b(z),z
       |p\alpha(1-\alpha)\bigr)\, .
  \label{3.27}
\end{eqnarray}
Here, the new path integrals are of the form
\begin{eqnarray}
  {\cal K}_0\bigl({\bf r}',z';{\bf r},z|\mu\bigr)
  &=& \frac{\mu}{2\pi\, i\, (z'-z)}
    \exp\left[ { {i\mu}
           \left({\bf r}' - {\bf r}\right)^2 \over {2\, (z'-z)} }
           \right]\, ,
  \label{3.28} \\
  {\cal K}\bigl({\bf r}',z';{\bf r},z|\mu\bigr)
  &=&   \int {\cal D}{\bf r}\, 
  \exp\left[  i\, \int\limits_z^{z'}\, d\xi\,
  \left[{\mu\over 2}\dot{\bf r}^2 
  + i\, \frac{1}{2} n(\xi)\, \sigma\left({\bf r}\right)
  \right] \right]\, .
  \label{3.29}
\end{eqnarray}
In the high-energy limit, this path-integral coincides 
with the average (\ref{2.26}) of two eikonal (fundamental) Wilson lines,
\begin{equation}
  \lim_{\mu \to \infty} {\cal K}\bigl({\bf r}',z';{\bf r},z|\mu\bigr)
  =  \exp\left[ - \frac{1}{2} n(\xi)\, \sigma\left({\bf r}\right)
  \right] \, \delta^{(2)}({\bf r}' - {\bf r})
  \, .
  \label{3.30}
\end{equation}
The path-integral (\ref{3.29}) determines the $O(1/E)$-corrections
to this eikonal average. Quite generally, it is seen to compare the
relative distance (\ref{3.26}) between the trajectories of the 
two generalized Wilson lines as a function of $\xi$. 

\subsection{The medium-induced gluon radiation spectrum}
\label{sec3c}
We consider now the medium-induced radiation amplitude for a quark 
of color $a$ splitting into a quark and a gluon of colors $b$ and 
$c$ respectively. The diagram for arbitrarily many  $M$-, $L$-, and 
$N$-fold scatterings of the in- and outgoing partons is\\
\begin{eqnarray}
&&\epsfxsize=8.0cm 
\hspace{-2cm}\centerline{\epsfbox{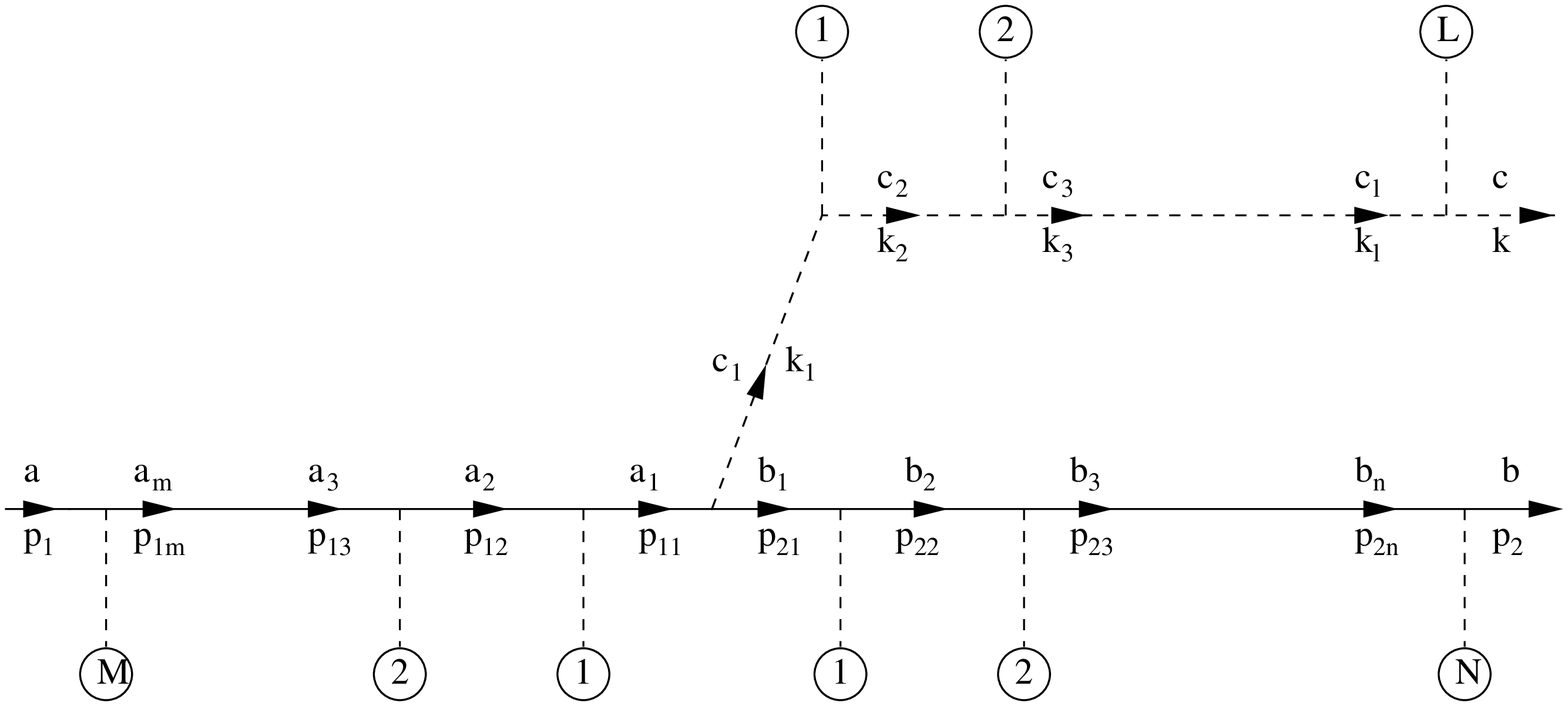}}
\vspace{-1cm}
\label{3.31} 
\end{eqnarray}
These multiple scattering contributions are summed up in the
Green's functions (\ref{3.9}) which include the $O(1/E)$ corrections
to the eikonal Wilson lines.  The corresponding radiation amplitude
reads
\begin{eqnarray}
  &&{\cal M}_{a\to b\, c} = -i \int dy_l\, 
  e^{i\,\bar{q}\, y_l}\, e^{-\epsilon\, |y_l|}
  \int d^2{\bf y} \int d^2{\bf x}_1\, d^2{\bf x}_2\, 
  d^2{\bf x}_g\, 
  \nonumber \\
  &&\qquad \times  
  e^{i{\bf x}_{1}\cdot {\bf p}_{1}
               - i{\bf x}_{2}\cdot {\bf p}_{2}
               - i{\bf x}_{g}\cdot {\bf k}}
  e^{-i\frac{{\bf p}_{1}^2}{2p_1}x_{l}
      +i\frac{{\bf p}_{2}^2}{2p_2}x_{l}
      -i\frac{{\bf k}^2}{2\omega}x_{l}}
  \nonumber \\
  &&\qquad \times G_{(q)}^{a a_1}({\bf x}_1;{\bf y}|p_1)\, 
  \hat{\Gamma}_{\bf y} \left(T^{c_1}\right)_{a_1b_1}\, 
  G_{(g)}^{c_1c}({\bf y};{\bf x}_g|\omega)\,
  G_{(q)}^{b_1b}({\bf y};{\bf x}_2|p_2)\, ,
  \label{3.32}
\end{eqnarray}
where the interaction vertex reads
\begin{equation}
  \hat{\Gamma}_{\bf y} = v^\dagger(p_1)\, \hat{D}_1 \gamma_0
  \epsilon\cdot \gamma \hat{D_2}\, v(p_2)\, .
  \label{3.33}
\end{equation}
Here, the differential operators $\hat{D}_i$ stem from the
Furry approximations of the quark lines, see eq. (\ref{3.15}).
The gluon radiation amplitude (\ref{3.32}) is a typical example
for which the order $O(E^0)$ term in the norm of the Green's function 
vanishes. As a consequence, the derivatives 
${\partial\over \partial {\bf y}}$ which are subleading in the 
Green's function itself, appear to leading
order in the transition amplitude and in the radiation cross section,
\begin{eqnarray}
  &&{d^3\sigma^{(in)}\over d(\ln x)\, d{\bf k}}
  = {\alpha_s\over (2\pi)^2}\, {1\over 4\, E_1^2\, (1-x)^2}
    \int d{\bf p}\, 
    \langle \vert{\cal M}_{a\to b\, c}\vert^2\rangle
    \nonumber\\
  &&= {\alpha_s\over (2\pi)^2}\, 
                {1\over x^2\, E_1^2\, (1-x)^2}
    2\, {\rm Re}\, 
    \int_{z_-}^{z_+} dy_l\, \int_{y_l}^{z_+} d{\bar y}_l\, 
    e^{i\bar{q}(y_l-{\bar y}_l) -\epsilon\, |y_l|-\epsilon\, |{\bar y}_l| }\, 
    \nonumber \\
    && \times 
    \int d{\bf y}\, d{\bf \bar y}\,
    d{\bf x}_1\, d{\bf \bar x}_1\,
    d{\bf x}_g\, d{\bf \bar x}_g\, d{\bf x}_2\,
    e^{- i{\bf k}\cdot ({\bf x}_g - {\bf \bar x}_g)}\, .
         \nonumber \\
    &&\times \Bigg \langle 
    G_{(q)}^{aa_1}({\bf x}_1;{\bf y}|p_1)\, T^{c_1}_{a_1b_1}\,
    \left({\partial\over \partial {\bf y}} 
           G_{(g)}^{c_1c}({\bf y};{\bf x}_g|\omega)\right)\, 
    G_{(q)}^{b_1b}({\bf y};{\bf x}_2|p_2)\, 
    \nonumber \\
    && \qquad  G_{(q)}^{b\bar{b}_1}({\bf x}_2;{\bf \bar y}|p_2)\,
    \left({\partial\over \partial {\bf \bar y}}  
           G_{(g)}^{c\bar{c}_1}({\bf \bar x}_g;{\bf \bar y}|\omega)\right)\,
    T^{{\bar c}_1}_{\bar{b}_1\bar{a}_1}\, 
    G_{(q)}^{\bar{a}_1a}({\bf \bar y};{\bf \bar x}_1|p_1)\Bigg \rangle \, .
    \label{3.34} 
\end{eqnarray}
Graphically, this can be visualized as
\begin{eqnarray}
&&\epsfxsize=7.0cm 
\hspace{-1cm}\centerline{\epsfbox{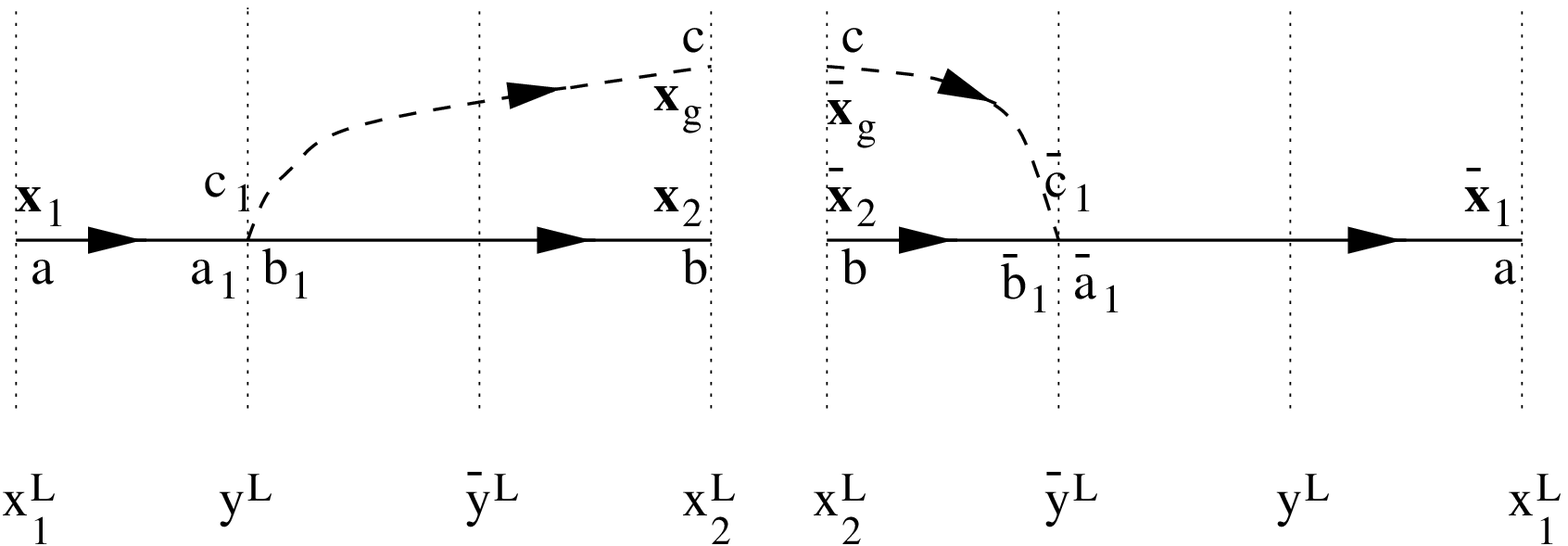}}
    \label{3.35} 
\end{eqnarray}
As discussed earlier, the softer parton, which in this case is the gluon, 
rescatters much more efficiently. Keeping only the leading $O(x)$ terms 
the final result 
for this radiation cross section reads\cite{Wiedemann:2000za}
\begin{eqnarray}
  &&{d^3\sigma^{(in)}\over d(\ln x)\, d{\bf k}}
  = {\alpha_s\over (2\pi)^2}\, {1\over \omega^2}\,
    C_F\, 
    2{\rm Re} \int_{z_-}^{z_+} dy_l  
  \int_{y_l}^{z_+} d\bar{y}_l\, 
  e^{-\epsilon |y_l|\, -\epsilon |\bar{y}_l|}
  \nonumber \\
  && \quad \times  \hspace{-0.1cm}
  \int \hspace{-0.1cm} d{\bf u}\,   e^{-i{\bf k}\cdot{\bf u}} 
  e^{ -\frac{1}{2} \int_{\bar{y}_l}^{z_+} d\xi\, n(\xi)\, 
    \sigma({\bf u}) }\,
    {\partial \over \partial {\bf y}}\cdot 
  {\partial \over \partial {\bf u}}\, 
  {\cal K}({\bf y}=0,y_l; {\bf u},\bar{y}_l|\omega) \, .
    \label{3.36}
\end{eqnarray}
The $\epsilon$-regularization in this expression has been introduced 
to suppress the contributions of gluons emitted at $t=\pm\infty$. 
With this adiabatic switching on (off) of the interaction at 
asymptotically early (late) times, the state of a single quark is 
an eigenstate of the Hamiltonian at $t\rightarrow\pm\infty$, and thus 
the medium induced emission of an extra gluon is indeed a transition 
to a different eigenstate. One could in principle perform the calculation 
without introduction of this explicit regulator. However in this case one 
must take care to calculate the transition amplitude between  eigenstates
of the interacting Hamiltonian and thus take into account $O(\alpha_s)$ 
corrections to the wave functions of the  initial and final states. 
The initial ``dressed quark'' state in such a calculation will contain 
the admixture of a $|qg\rangle$ component as discussed in Section 2. 
To calculate 
the radiation spectrum one then needs to subtract from the final state 
wave function the component which overlaps with the ``dressed quark'' wave 
function. Introduction of the regulator makes the calculation technically 
much simpler. However, care is needed since the limit $\epsilon \to 0$ 
in (\ref{3.36}) does not commute with the longitudinal integrations.
In practice, one can remove the regularization e.g. by splitting
up the longitudinal $y_L$- and $\bar{y}_L$-integrations 
of the radiation cross sections into six parts,
\begin{eqnarray}
  \int\limits_{z_-}^{z_+} \int\limits_{y_L}^{z_+}
  &=& \int\limits_{z_-}^{0} \int\limits_{y_L}^{0}
  + \int\limits_{z_-}^{0} \int\limits_{0}^{L}
  + \int\limits_{z_-}^{0} \int\limits_{L}^{z_+}
  + \int\limits_{0}^{L} \int\limits_{y_L}^{L}
  + \int\limits_{0}^{L} \int\limits_{L}^{z_+}
  + \int\limits_{L}^{z_+} \int\limits_{y_L}^{z_+}\, ,
  \label{3.37}
\end{eqnarray}
and to write the radiation cross section as a sum of the 
corresponding six contributions
\begin{eqnarray}
 &&\frac{d^3\sigma}{d({\rm ln}x)\,d{\bf k}}
   =
 \frac{\alpha_s}{\pi^2}\, C_F\, 
 \left(I_1 + I_2 + I_3 + I_4 + I_5 + I_6 \right)\, . 
 \label{3.38}
\end{eqnarray}
This allows to remove in all terms $I_j$ the $\epsilon$-regularization 
of (\ref{3.36}). In what follows, we are interested in the
gluon energy distribution of a quark produced in the medium
at position $z_- = 0$. In this case, the contributions $I_1$,
$I_2$, $I_3$ to (\ref{3.37}) vanish. The remaining three terms
denote the {\it direct production term} $I_4$ for gluon emission
inside the target in both amplitude and complex conjugate 
amplitude, the {\it destructive interference term} $I_5$ which is the
characteristic interference term between gluon emission in and 
beyond the target, and the {\it hard vacuum radiation term} $I_6$,
\begin{equation}
\epsfxsize=7.0cm 
\centerline{\epsfbox{ii456pap.epsi}}
\label{3.39}
\end{equation}
The contribution $I_6$ can be 
associated to the medium-independent contribution to the
hard radiation off a nascent quark jet propagating in the 
vacuum \cite{Wiedemann:2000za}. This is a consequence of the leading $O(x)$
approximation in which only the rescattering of the radiated gluon 
contributes, and the rescattering of the hard quark in (\ref{3.39})
is neglected,
\begin{equation}
  I_6 = \frac{1}{{\bf k}^2}\, .
  \label{3.40}
\end{equation}
This term may be viewed as the medium-unaffected DGLAP part of the gluon 
radiation. In the following, we shall discuss the 
medium-induced modification of the gluon radiation spectrum,
determined by $I_4$ and $I_5$. For notational convenience, we
shall not denote explicitly that the medium-independent term 
$I_6$ is subtracted.

\section{Properties of the medium-induced gluon radiation spectrum}
\label{sec4}
Two approximation schemes for the radiation cross section (\ref{3.36})
have been explored recently: the dipole 
approximation~\cite{Baier:1996kr,Zakharov:1996fv,Baier:1998kq,Wiedemann:2000za}
and the opacity expansion\cite{Wiedemann:2000za,Gyulassy:2000er}. 
In the dipole approximation, the 
medium-dependence results from the multiple soft scattering 
of the projectile in the spatially extended matter. The
opacity expansion can be related to an expansion in the
effective number of scatterings and allows for hard
momentum transfers from the medium. Here, we discuss for
both approximations the inclusive energy 
distribution of gluon radiation off an in-medium 
produced parton. Integrating over the transverse momentum
of the radiated gluon up to $\vert {\bf k}\vert < \chi\, \omega$,
we have~\cite{Wiedemann:2000za,Wiedemann:2000tf}
\begin{eqnarray}
  \omega\frac{dI}{d\omega}
  &=& {\alpha_s\,  C_R\over (2\pi)^2\, \omega^2}\,
    2{\rm Re} \int_{\xi_0}^{\infty}\hspace{-0.3cm} dy_l
  \int_{y_l}^{\infty} \hspace{-0.3cm} d\bar{y}_l\,
   \int d^2{\bf u}\,  \int_0^{\chi \omega}\, d^2{\bf k}\, 
  e^{-i{\bf k}\cdot{\bf u}}   \,
  e^{ -\frac{1}{2} \int_{\bar{y}_l}^{\infty} d\xi\, n(\xi)\,
    \sigma({\bf u}) }\,
  \nonumber \\
  && \times {\partial \over \partial {\bf y}}\cdot
  {\partial \over \partial {\bf u}}\,
  \int_{{\bf y}=0}^{{\bf u}={\bf r}(\bar{y}_l)}
  \hspace{-0.5cm} {\cal D}{\bf r}
   \exp\left[ i \int_{y_l}^{\bar{y}_l} \hspace{-0.2cm} d\xi
        \frac{\omega}{2} \left(\dot{\bf r}^2
          - \frac{n(\xi) \sigma\left({\bf r}\right)}{i\, \omega} \right)
                      \right]\, .
    \label{4.1}
\end{eqnarray}
The radiation off hard quarks or gluons differs by 
the Casimir factor $C_R = C_F$ or $C_A$, respectively. The 
properties of the medium enter eq. (\ref{4.1}) by the product of
the density $n(\xi)$ of scattering centers times 
the dipole cross section $\sigma({\bf r})$ which determines 
the strength of a single elastic scattering, see (\ref{2.14}).
Note that the density can be time-dependent as is for example 
the case in expanding medium.

\subsection{Multiple soft scattering}
\label{sec4a}
Consider a medium-dependence introduced by
arbitrary many soft scattering centers, rather than a few hard
ones. 
In such a medium the projectile performs a Brownian motion in transverse 
plane; the transverse position of the projectile
in configuration space changes rather smoothly and small 
relative distances in the dipole cross section are 
important. The integrand of
(\ref{3.36}) has its main support at small transverse distances
$r = |{\bf r}|$ which allows to write the dipole cross section
to logarithmic accuracy as~\cite{Zakharov:1996fv,Zakharov:1998sv}
\begin{eqnarray}
  n(\xi)\, \sigma({\bf r}) \simeq \frac{1}{2}\, \hat{q}(\xi)\, {\bf r}^2\, .
  \label{4.2}
\end{eqnarray}
Here, $\hat{q}(\xi)$ is the transport coefficient\cite{Baier:1996sk} 
which characterizes the transverse momentum squared $\mu^2$ transferred 
to the projectile per mean free path $\lambda$. For a static medium, 
it is time-independent,
\begin{equation}
  \hat{q} = \frac{\mu^2}{\lambda}\, .
  \label{4.3}
\end{equation}
In the multiple soft scattering approximation, the transport
coefficient $\hat{q}$ and the in-medium pathlength $L$ are
the only informations about the properties of the medium
which enter the gluon radiation spectrum (\ref{4.1}).

\subsubsection{The harmonic oscillator approximation (static medium)}
\label{sec4a1}
In the so-called dipole approximation (\ref{4.2}), the
path integral (\ref{3.29}) in the gluon radiation cross section 
(\ref{3.36}) is equivalent to that of a harmonic 
oscillator\cite{Zakharov:1998sv,Wiedemann:2000tf}
\begin{eqnarray}
  {\cal K}_{\rm osc}\bigl({\bf y},y_l;{\bf u},\bar{y}_l|\mu\bigr) 
     &=& {A\over \pi\, i} \exp\left[iAB({\bf y}^2 + {\bf u}^2)
             -2\,i\,A\,{\bf y}\cdot{\bf u} \right]\, ,
  \label{4.4} \\
  A &=&  {\mu\Omega\over 2\, \sin(\Omega\, \Delta y)}\, ,\qquad
  B = \cos(\Omega\, \Delta y)\, ,
  \label{4.5}
\end{eqnarray}  
with complex oscillator frequency
\begin{equation}
  \Omega = \frac{1-i}{\sqrt{2}}\, \sqrt{\hat{q}\over 2 \omega}\, .
  \label{4.6}
\end{equation}
\underline{Explicit solution for parton produced inside the medium:}\\
Inserting ${\cal K}_{\rm osc}$ into $\omega \frac{dI}{d\omega}$,
the medium-induced gluon energy distribution becomes
\begin{eqnarray}
 \omega \frac{dI}{d\omega} &=& \frac{\alpha_s}{\pi^2}\, C_R\, 
 \left(\bar{I}_4 + \bar{I}_5 \right)\, ,
 \label{4.7}\\
  \bar{I}_j &\equiv& \int_0^{2\pi} d\varphi_k 
              \int_0^{\chi\, \omega} k_\perp\, dk_\perp\, I_j\, ,
  \label{4.8}
\end{eqnarray}
where the terms $I_j$ are defined by eq. (\ref{3.39}). Their
explicit form is 
\begin{eqnarray}
  \bar{I}_4 &=& 4\pi {\rm Re}  \int_0^{\tilde{L}} d\tilde{z}_1
                \int_0^{\tilde{z}_1} d\tilde{z}_2
                \left( \frac{i}{2} \frac{1}{1-\cosh[(1+i)\Delta \tilde{z}]}
                \right.
                \nonumber \\
            &&  \qquad \qquad + \left. 
                \frac{i}{4} \, 
                e^{- 2M_c^2 \sinh[(1+i)\Delta \tilde{z}/2] / 2}
                \, \frac{F}{N^2}
                \right)\, ,
                \label{4.9} \\
  \bar{I}_5 &=& 4\pi {\rm Re}  \int_0^{\tilde{L}} d\tilde{z}
                \left( \frac{-1}{2} \frac{1+i}{\sinh[(1+i)\tilde{z}]}
                  \right.
                  \nonumber \\
                && \qquad \qquad + \left.
                \frac{(1+i)\, e^{-(1+i)M_c^2 \tanh[(1+i)\tilde{z}/2]}}
              {4\, \sinh[(1+i)\tilde{z}/2] \cosh[(1+i)\tilde{z}/2]}\right)\, ,
            \label{4.10}\\
           F  &\equiv& 
                \tilde{z}_2^2 - 2i\coth[(1+i)\Delta \tilde{z}/2]
                \left[ (1+i)(M_c^2+\tilde z_2) 
                  \right.
                \nonumber \\
                && \qquad \qquad
                + \left. \coth[(1+i)\Delta \tilde{z}/2]\right]\, ,
                \label{4.11}\\
  N &\equiv& \tilde{z}_2\, 
        \sinh\left( \frac{1+i}{2}(\tilde{z}_1-\tilde{z}_2)\right)
         + (1-i)\, 
        \cosh\left( \frac{1+i}{2}(\tilde{z}_1-\tilde{z}_2)\right)\, .
  \label{4.12}
\end{eqnarray}
Here, the longitudinal distances are written in dimensionless units,
rescaled by the modulus of the oscillator frequency $\vert\Omega\vert$,
\begin{eqnarray}
 \tilde{z} = \sqrt{2} |\Omega|\, z\, ,\qquad
 \tilde{L} = \sqrt{2} |\Omega|\, L\, .
 \label{4.13}
\end{eqnarray}
The upper integration limit of the $k_\perp$ integral takes in
dimensionless variables the from
\begin{equation}
  M_c = \frac{\chi\, \omega}{\sqrt{ 2\omega \sqrt{2} |\Omega|}}\, .
  \label{4.14}
\end{equation}
\underline{Totally coherent limit for parton in initial state:}\\
Assume that a parton is part of an initial state. In this case,
prior to scattering on a target, the parton has already a fully 
developped wavefunction with additional virtual gluons. In the
scattering process, these virtual gluons can be freed. As a consequence,
the gluon energy distribution radiated off this parton has to 
include radiation vertices at arbitrary early times. One needs
to include all six terms $I_j$ of (\ref{3.38}) in calculating
eq. (\ref{4.7}). Explicit expressions for this case are given 
in Ref.~\cite{Wiedemann:2000tf}.
Here, we consider only the high-energy limit of the resulting
expression. Since $\Omega \propto \frac{1}{\sqrt{\omega}}$, 
this high-energy limit of the radiation spectrum corresponds 
to the leading order in $O(\tilde{L} = \sqrt{2} |\Omega|\, L)$.
It reads
\begin{eqnarray}
 &&\omega \frac{d^3I^{(in)}}{d\omega\,d{\bf k}} =
 \frac{\alpha_s}{\pi^2} 
          {4\pi C_R  \over \hat{q}\, L}
 \int \frac{d^2{\bf q}}{(2\pi)^2}\, 
          e^{-\frac{{\bf q}^2}{\hat{q}\, L}}\, 
          \frac{{\bf q}^2}
                {({\bf q}-{\bf k})^2\, {\bf k}^2}\, .
  \label{4.15}
\end{eqnarray}
This is the totally coherent limit of the gluon radiation spectrum
for an incoming parton. It is the Gunion-Bertsch gluon radiation 
spectrum (\ref{2.28})
for a target momentum transfer ${\bf q}$ which is acquired 
by the transverse Brownian motion of the incoming projectile,

\subsubsection{Qualitative estimates of $\omega \frac{dI}{d\omega}$ vs.
quantitative calculations}
\label{sec4a2}

Important properties of the gluon energy distribution can be
obtained from simple estimates. Their range of applicability, however,
depends on phase space constraints which are not as easy to estimate.
Here, we compare qualitative arguments to numerical 
calculations\cite{Wiedemann:2000tf,Salgado:2003gb}.

{\it Qualitative arguments:}\cite{Baier:2002tc}
We consider a gluon in the hard parton 
wave function. This gluon is emitted due to multiple scattering
if it picks up sufficient transverse momentum to decohere from the partonic
projectile. For this, the average phase $\varphi$ accumulated by
the gluon should be of order one,
\begin{equation}
  \varphi = \Bigg\langle \frac{k_\perp^2}{2\omega}\, \Delta z \Bigg\rangle
  \sim \frac{\hat{q}\, L}{2\omega} L = \frac{\omega_c}{\omega}\, .
  \label{4.16}
\end{equation}
Thus, for a hard parton traversing a finite pathlenth $L$ in the medium,
the scale of the radiated energy distribution is set by 
the ``characteristic gluon frequency''
\begin{equation}
  \omega_c = \frac{1}{2}\, \hat{q}\, L^2\, .
  \label{4.17}
\end{equation}
For an estimate of the shape of the energy distribution, we 
consider the number $N_{\rm coh}$ of scattering centers which 
add coherently in the gluon phase (\ref{4.16}), 
$k_\perp^2 \simeq N_{\rm coh}\, \mu^2$. Based on expressions
for the coherence time of the emitted gluon, 
$t_{\rm coh} \simeq \frac{\omega}{k_\perp^2} \simeq 
\sqrt{\frac{\omega}{\hat{q}}}$
and $N_{\rm coh} = \frac{t_{\rm coh}}{\lambda} = 
\sqrt{\frac{\omega}{\mu^2\, \lambda}}$, one estimates for the
gluon energy spectrum per unit pathlength
\begin{equation}
  \omega \frac{dI}{d\omega\, dz} \simeq 
  \frac{1}{N_{\rm coh}}\, 
  \omega \frac{dI^{\rm 1\, scatt}}{d\omega\, dz} \simeq
  \frac{\alpha_s}{t_{\rm coh}}
  \simeq \alpha_s\, \sqrt{\frac{\hat{q}}{\omega}}\, .
  \label{4.18}
\end{equation}
This $1/\sqrt{\omega}$-energy dependence of the
medium-induced non-abelian gluon energy spectrum is
expected for sufficiently small $\omega < \omega_c$.

{\it Quantitative analysis:}
The gluon energy distribution
(\ref{4.1}) depends not only on $\omega_c$, but also on
the dimensionless ``density parameter''\cite{Salgado:2002cd}
\begin{equation}
  R= \frac{1}{2}\, \hat{q}\, \chi^2\, L^3\, .
  \label{4.19}
\end{equation}
A finite value of $R$ implements the kinematical constraint
$k_\perp < \chi \omega$ in the transverse momentum phase space 
of the emitted gluon, see (\ref{4.1}). The kinematical boundary is $\chi = 1$. 
This constraint is neglected in the argument leading to 
the $1/\sqrt{\omega}$-energy dependence of (\ref{4.18}). 

%
\begin{figure}[t]\epsfxsize=10.0cm 
\centerline{\epsfbox{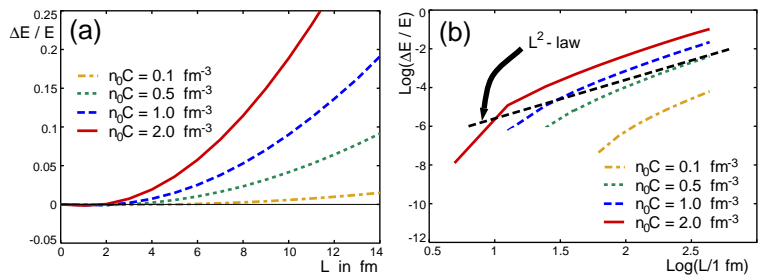}}
\vspace{0.5cm}
\caption{(a) Dependence of the medium-induced radiative energy loss 
$\langle \Delta E\rangle$ on the in-medium pathlength $L$ for $E = 100$ GeV.
The transport coefficient is defined here as $\frac{\hat{q}}{2} = n_0\, C$.
(b) Double logarithmic presentation of (a).
Figure taken from \protect\cite{Wiedemann:2000tf}.
}\label{fig1}
\end{figure}
%
The limit $R \to \infty$ removes the kinematical constraint
from (\ref{4.1}) and coincides with the original result of Baier, 
Dokshitzer, Mueller, Peign\'e and Schiff\cite{Baier:1996sk},
\begin{equation}
  \lim_{R\to \infty}\, 
   \omega \frac{dI}{d\omega} =
   \frac{2\alpha_s C_R}{\pi}\, 
   \ln \Bigg \vert
   {\cos\left[\,(1+i)\sqrt{\frac{\omega_c}{2\omega}}\,\right]}
   \Bigg \vert \, .
   \label{4.20}
\end{equation}
The limiting case for small $\omega$ is as expected from
the estimates (\ref{4.18}), \cite{Baier:2001yt}
\begin{eqnarray}
   \lim_{R\to \infty}\, 
   \omega \frac{dI}{d\omega} \simeq 
           \frac{2\alpha_s C_R}{\pi} 
          \left\{ \begin{array} 
                  {r@{\qquad  \hbox{for}\quad}l}                 
                  \sqrt{\frac{\omega_c}{2\, \omega}}
                  & \omega < \omega_c\, , \\ 
                  \frac{1}{12} 
                  \left(\frac{\omega_c}{\omega}\right)^2
                  & \omega > \omega_c\, . 
                  \end{array} \right.
  \label{4.21}
\end{eqnarray}
The average parton energy loss is the zeroth moment of this
energy distribution
\begin{equation}
 \langle \Delta E \rangle_{R\to\infty} = 
 \lim_{R\to \infty}\,  \int d\omega\, 
   \omega \frac{dI}{d\omega}
 \simeq   \frac{\alpha_s C_R}{2}\, \omega_c
 \propto L^2\, . 
 \label{4.22}
\end{equation}
This is the well-known $L^2$-dependence of the average energy 
loss~\cite{Baier:1996kr,Baier:1996sk,Zakharov:1997uu}. The
numerical calculation in Figure ~\ref{fig1} shows deviations
from the limiting case (\ref{4.22}). [In Ref.\cite{Wiedemann:2000tf},
an erroneous prefactor $N_c$ was included in the definition of
$\omega \frac{dI}{d\omega}$; hence, the values for $\Delta E$ shown in 
Fig.\ref{fig1} and \ref{fig3} are a factor 3 too large.]
In general, the
strong increase of the average energy loss $\Delta E$ with
in-medium pathlength indicates the strong sensitivity of
partonic energy loss to the geometry of the nuclear collision.
%
\begin{figure}[h]\epsfxsize=10.7cm
\centerline{\epsfbox{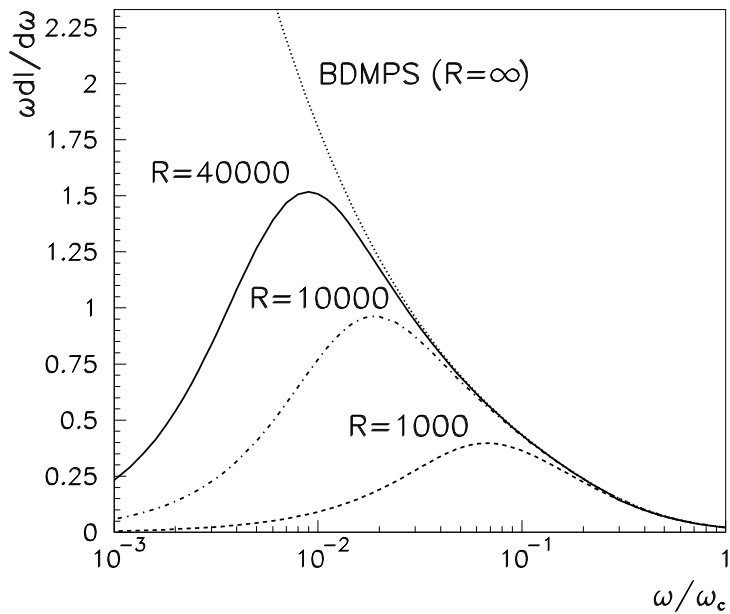}}
\caption{The energy distribution of radiated gluons
$\omega \frac{dI}{d\omega}$ for different values of the 
kinematical constraint $R = \omega_c\, L$. Figure taken
from \protect\cite{Salgado:2003gb}.
}\label{fig2}
\end{figure}

The average energy loss $\Delta E$ is the first moment of
the energy distribution (\ref{4.1}). Fig.~\ref{fig2} shows
this distribution for finite values of the density parameter $R$. 
At sufficiently large gluon energy, the distribution approaches 
for any value of $R$ the BDMPS limit (\ref{4.20}) . Below
a critical gluon energy $\bar{\omega}$, however, the finite size
gluon spectrum is depleted in comparison to the BDMPS limit.
This suppression results from a phase space constraint:
gluons are radiated on average at a characteristic angle
\cite{Salgado:2003gb},
\begin{equation}
  \Theta_c^2 \simeq \frac{k_\perp^2}{\omega^2}
  \simeq \frac{\sqrt{\omega \hat{q}}}{\omega^2}
  \simeq \left( \frac{\omega}{\omega_c}\right)^{-3/2}
  \frac{1}{R} < 1\, .
  \label{4.23}
\end{equation}
This depletes the gluon distribution for gluon energy 
$\bar{\omega}$ below 
\begin{equation}
  \frac{\bar{\omega}}{\omega_c} \propto 
        \left(\frac{1}{R} \right)^{2/3}\, .
  \label{4.24}
\end{equation}
This $R$-dependent suppression is seen clearly in 
Fig.~\ref{fig2}. 

\subsubsection{Angular Dependence of the gluon energy distribution}
\label{sec4a3}
Quantum Chromodynamics is a finite resolution theory;
the energy of a single quark or gluon is not a measurable 
quantity. One rather measures the {\it fraction} of the hadronic 
decay products of a single quark or gluon which are radiated within 
a certain jet opening angle. Thus, the angular dependence of the
medium-induced modification of gluon bremsstrahlung is important.

The angular dependence of 
$\omega \frac{dI}{d\omega}$ can be studied by varying the kinematic
cut-off $\chi\, \omega$ of the transverse momentum 
integration \cite{Baier:1999ds,Wiedemann:2000tf,Baier:2001qw}.
For fixed characteristic gluon energy 
$\omega_c = \frac{1}{2} \hat{q} L^2$ and fixed dimensionless
parameter $\frac{1}{2} \hat{q} L^3$, the radiation
spectrum $\omega \frac{dI}{d\omega}$ {\it inside} a given angle 
$\Theta=\arcsin(\chi)$ depends on the parameters
$\omega_c$ and $R = \chi^2  \frac{1}{2} \hat{q} L^3$. From
Fig.~\ref{fig2} one thus concludes immediately that the
harder gluons are emitted more collinear. 

From the gluon energy distribution inside a given cone opening
angle, one obtains the medium-induced additional average 
energy emitted {\it outside} a cone of given opening angle $\Theta$, 
\begin{eqnarray}
        \frac{\Delta E}{E}(\Theta=\arcsin(\bar\chi)) = 
        \frac{\Delta E}{E}(\chi=1) - \frac{\Delta E}{E}(\bar{\chi})
        \label{4.25}\, .
 \end{eqnarray}
%
\begin{figure}[h]\epsfxsize=10.0cm 
\centerline{\epsfbox{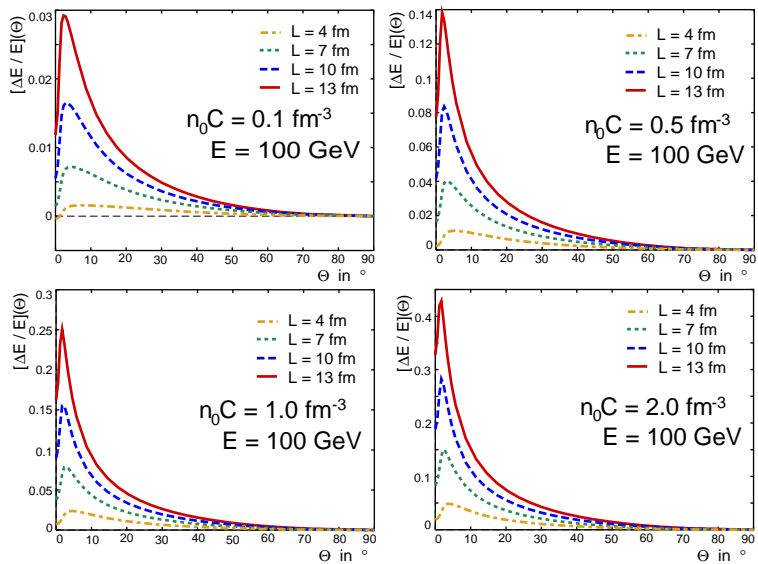}}
\vspace{0.5cm}
\caption{The fraction 
of the total 
radiative energy loss $\Delta E/E$ emitted outside a jet cone of fixed angle 
$\Theta$. Here, the transport coefficient is defined as 
$\hat{q} = 2\, n_0\, C$. Figure taken from 
\protect\cite{Wiedemann:2000tf}.
}\label{fig3}
\end{figure}
%
As seen from Fig.~\ref{fig3}, $\frac{\Delta E}{E}(\Theta)$ does
not decrease monotonously with increasing $\Theta$ but has a
maximum at finite jet opening angle. The reason is that the
radiative energy loss outside a cone angle $\Theta$ receives
additional contributions from the Brownian $k_\perp$-broadening
\begin{equation}
        \frac{1}{{\bf k}^2} \longrightarrow
         \frac{1}{({\bf k}+ {\bf q})^2}
        \label{4.26}
\end{equation}
of the hard vacuum radiation term $I_6$.
Such contributions do not affect the total ${\bf k}$-integrated 
yield $ \frac{\Delta E}{E}(\Theta=0)$, since they result only in 
a shifting of the transverse momentum phase space distribution of
the emitted gluon. However, this shift in transverse phase space
shows up as soon as a finite cone size is chosen. We conclude from 
Fig.~\ref{fig3} that the total ${\bf k}$-integrated radiative 
energy loss$ \frac{\Delta E}{E}(\Theta=0)$
is not the upper bound for the radiative energy loss outside a
finite jet cone angle $ \frac{\Delta E}{E}(\Theta)$. 
Depending on the transport coefficient and the in-medium path length, 
the latter can be larger by more than a factor $2$.

%
\subsubsection{Harmonic oscillator approximation (expanding medium)}
\label{sec4a4}
In nucleus-nucleus collisions at collider energies, the produced
hard partons propagate through a rapidly expanding medium. The
density of scattering centers and thus the transport coefficient
$\hat{q}(\xi)$ is expected to reach a 
maximal value $\hat{q}_d$ around the plasma formation time $\xi_0$,
and then decreases rapidly due to the strong longitudinal and 
- to a lesser extent - transverse expansion, 
\begin{equation}
  \hat{q}(\xi) = \hat{q}_d \left( \frac{\xi_0}{\xi} \right)^\alpha\, .
  \label{4.27}
\end{equation}
Here, $\alpha = 0$ characterizes the static medium discussed
above. The value $\alpha =1$ corresponds to a one-dimensional, 
boost-invariant longitudinal expansion and approximates the 
findings of hydrodynamical simulations. The formation time $\xi_0$ 
of the medium may be set by
the inverse of the saturation scale $p_{\rm sat}$~\cite{Eskola:1999fc} and 
is $\approx$ 0.2 fm/c at RHIC and $\approx$ 0.1 fm/c at LHC. Since the
time difference between the formation of the hard parton and the
formation of the medium bulk is irrelevant for the evaluation of
the radiation spectrum (\ref{4.1}), one can replace in (\ref{4.1}) 
the production time $\xi_0$ of the parton by $0$.
 
For a dynamically evolving medium of the type (\ref{4.27}),
the path-integral (\ref{3.29}) in (\ref{4.1}) is the path integral of 
a 2-dimensional harmonic oscillator with time dependent (imaginary) 
frequency $\Omega^2 (\xi)\equiv {\hat{q}(\xi)\over i2\omega}$
and {\it mass} $\omega$. The explicit solution can be written
in terms of the variables $z(\xi) = 2 i\nu\Omega(\xi_0)\xi_0
\left({\xi\over\xi_0}\right)^{1/2\nu}$ and 
$\nu = {1\over 2-\alpha}$: \cite{Baier:1998yf}
\begin{eqnarray}
\label{4.28}
  &&{\cal K}({\bf r}(y_l),y_l;{\bf r}(\bar{y}_l),\bar{y}_l|\omega) =\\
  &&{\omega\over 2\pi iD(\bar y_l,y_l)}\exp\left[-{\omega\over 
  2iD(\bar y_l,y_l)}\left(c_1 {\bf r}(\bar y_l)^2+
  c_2{\bf r}(y_l)^2-2 {\bf r}(\bar y_l)\cdot {\bf r}(y_l)\right)\right]\, .
  \nonumber
\end{eqnarray}
Here, the coefficients
are given by modified Bessel functions  $I_\nu(z)$ and $K_\nu(z)$ 
[here: $z \equiv z(y_l)$, 
$\bar{z} \equiv z(\bar{y}_L)$]
\begin{eqnarray}
  c_1 &\equiv& c_1(\bar{y}_L,y_L) = z(\bar{z}/z)^\nu\,  
  \left[I_{\nu-1}(z)K_\nu(\bar{z})+K_{\nu-1}(z)I_\nu(\bar{z})\right]\, ,
  \label{4.29}\\ 
  c_2 &\equiv& c_2(\bar{y}_L,y_L) = \bar{z} (z/\bar{z})^\nu\, 
  \left[K_\nu(z)I_{\nu-1}(\bar{z}) + I_\nu(z)K_{\nu-1}(\bar{z})\right]\, .
  \label{4.30}
\end{eqnarray}
and the fluctuation determinant reads
\begin{equation}
D(t,t_0)={2\nu\xi_0\over(2i\nu\Omega(t_o)\xi_0)^{2\nu}}(z z_0)^\nu
\left[I_\nu(z)K_\nu(z_0)-K_\nu(z)I_\nu(z_0)\right]\, .
\label{4.31}
\end{equation}
One checks that for $\alpha = 0$, $\nu = \frac{1}{2}$, the solution
(\ref{4.4}) is regained.
%
\begin{figure}[t]\epsfxsize=12.7cm
\centerline{\epsfbox{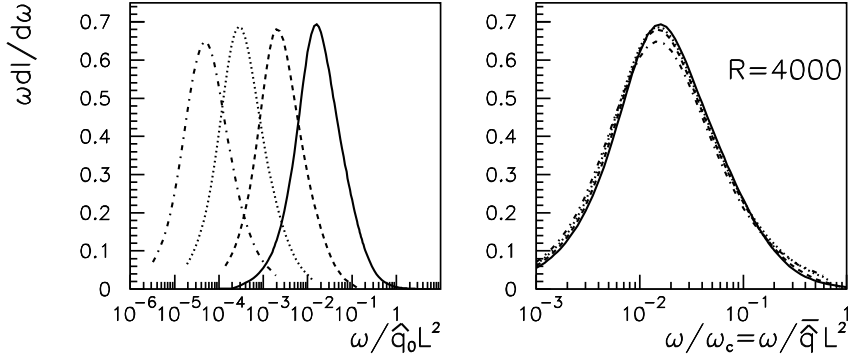}}
\caption{LHS: The medium-induced gluon energy distribution 
radiation for a dynamically expanding collision regions 
(\protect\ref{4.27}) with expansion parameter $\alpha = $
0, 0.5, 1.0 and 1.5. The value of the transport coefficient
$\hat{q}_0$ is taken at initial time $\xi_0$.
RHS: The same gluon radiation spectrum with parameters
rescaled according to (\ref{4.32}). Figure taken from
\protect\cite{Salgado:2003gb}.
}\label{fig4}
\end{figure}

Using the explicit form of the path integral (\ref{4.28}), one
can calculate the medium-induced gluon energy distribution (\ref{4.1})
for a dynamically expanding medium~\cite{Baier:1998yf}. The result is 
shown in Fig.\ref{fig4}. The radiation spectrum 
$\omega \frac{dI}{d\omega}$ satisfies a simple scaling law which 
relates the radiation spectrum of a dynamically expanding
collision region to an equivalent static scenario. The 
linearly weighed line integral \cite{Salgado:2002cd}
\begin{equation}
  \overline{\hat{q}} = \frac{2}{L^2}\int_{\xi_0}^{\xi_0+L} d\xi\, 
  \left( \xi - \xi_0\right)\, \hat{q}(\xi) 
  \label{4.32}
\end{equation}
defines the transport coefficient of the equivalent static
scenario. 
The linear weight in (\ref{4.32}) implies that scattering centers
which are further separated from the production point of the
hard parton are more effective in leading to partonic energy
loss. In contrast to earlier believe that parton energy loss
is most sensitive to the hottest and densest initial stage of
the collision, this implies for a dynamical expansion following
Bjorken scaling [$\alpha = 1$ in eq. (\ref{4.27})] that all 
timescales contribute equally to the average transport coefficient. 
This makes partonic energy loss a valuable tool for the measurement
of the quark-gluon plasma lifetime.

\subsubsection{Properties of the transport coefficient}
\label{sec4a5}
The only property of the medium which enters the medium-induced
energy distribution (\ref{4.1}) is the transport coefficient 
\begin{equation}  
 \hat{q} = \mu^2/\lambda 
  = \int \frac{d{\bf q}}{(2\pi)^2}\, {\bf q}^2\,  
                     |a_+({\bf q})|^2 \,
  \label{4.33}
\end{equation}
For phenomenological estimates of the value $\hat{q}$ for 
{\it cold} nuclear matter, one often invokes the relation of 
$\hat{q}$ to the gluon structure function 
$\hat{q} = \frac{4\pi^2\, \alpha_s N_c}{N_c^2 - 1} \rho 
\left[ xG(x,\hat{q}L)\right]$. 
This leads to estimates~\cite{Baier:1996sk} 
$\hat{q}_{\rm cold} < (200\, {\rm MeV})^2/ {\rm fm}$.
There are alternative approaches~\cite{Baier:1996sk,Arleo:2002ph} 
which extract $\hat{q}_{\rm cold}$ from the 
medium modification of hard processes in ``cold nuclear matter'', 
such as medium-modifications to the Drell-Yan 
production \cite{Arleo:2002ph} in $p-A$ or the dijet imbalance 
in $e-A$~\cite{Luo:ui,Luo:np}.   
The extracted values are consistent with 
$\hat{q}_{\rm cold} < (200\, {\rm MeV})^2/ {\rm fm}$,
but the systematic uncertainties of these determinations are 
difficult to specify.
%
\begin{figure}[t]\epsfxsize=7.7cm
\centerline{\epsfbox{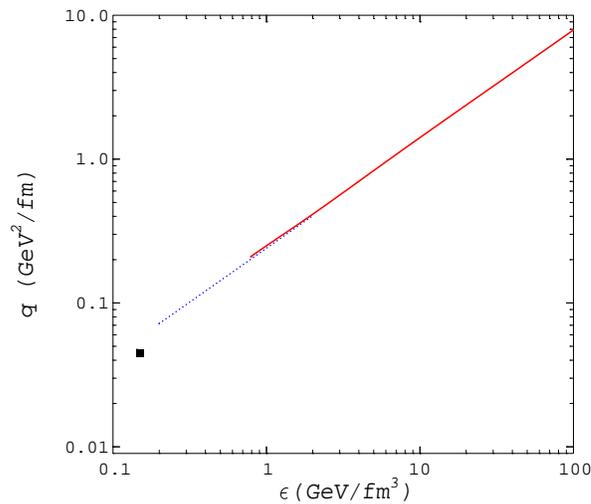}}
\caption{Estimate of the transport coefficient as a function of the
energy density $\epsilon$. Figure taken from 
\protect\cite{Baier:2002tc}.
}\label{fig5}
\end{figure}

For the transport coefficient in hot quark-gluon matter, even less is known. 
Baier, Dokshitzer, Mueller and Schiff~\cite{Baier:1996sk} estimated 
for the rescattering properties of hot matter at a temperature $T = 250$ MeV
a value $\hat{q} = (500\, {\rm MeV})^2/ {\rm fm}$. In any case, the transport
coefficient is expected to increase linearly with the energy density, as
seen in Fig.~\ref{fig5}. In the case of a phase transition at $T_c \approx
175$ MeV, this indicates a strong jump of $\hat{q}$ at $T_c$. The 
medium modification of hard processes observed at RHIC and LHC should allow 
to determine the value of $\hat{q}$ and thus provide an independent estimate
of the energy density attained in the collision. For this,
the proper treatment of the dynamical expansion of the 
collision region e.g. via (\ref{4.32}) is obviously important.

The transport coefficient has been related also to 
the saturation momentum scale\cite{Baier:2002tc}
\begin{equation}  
  Q_s^2 \simeq \hat{q}\, L\, . 
  \label{4.34}
\end{equation}
The saturation scale entering here should be defined for a color
octet dipole -- thus, it is a factor $C_A/C_F$ larger than the
saturation scale determined from the $q\bar{q}$ scattering
probability in DIS. The numerical value of $Q_s$ is very uncertain but 
$Q_s^2 \leq (3\, {\rm GeV})^2$ may be considered as an upper bound at LHC.
For in-medium path lengths $L$ up to twice the 
nuclear size, this corresponds to values up to $R = \frac{1}{2} \hat{q}
L^3 < 40000$. 

\subsection{Opacity Expansion of the radiation cross section}
\label{sec4b}

In what follows, we discuss the opacity expansion of the radiation
cross section (\ref{4.1}). This is based on the expansion of the 
path-integral (\ref{3.29}) in powers of the dipole cross section,
\begin{eqnarray}
 &&{\cal K}({\bf r},y_l;{\bf \bar r},\bar{y}_l) =
    {\cal K}_0({\bf r},y_l;{\bf \bar r},\bar{y}_l)
 \nonumber \\
 && - \int\limits_{z}^{z'}\, {\it d}\xi\, n(\xi)\, 
 \int {\it d}{\bf r}_1\,
 {\cal K}_0({\bf r},y_l;{\bf r}_1,\xi)\, \frac{1}{2}\,
   \sigma({\bf r}_1)\, 
   {\cal K}_0({\bf r}_1,\xi;{\bf \bar r},\bar{y}_l)
 \nonumber \\
 && + \int\limits_{z}^{z'} {\it d}\xi_1\, n(\xi_1)\, 
    \int\limits_{\xi_1}^{z'} {\it d}\xi_2\, n(\xi_2)\, 
    \int {\it d}{\bf r}_1\,{\it d}{\bf r}_2\,
    {\cal K}_0({\bf r},y_l;{\bf r}_1,\xi_1)\,  
    \nonumber \\
 && \times \frac{1}{2}\,
    \sigma({\bf r}_1)\, 
    {\cal K}({\bf r}_1,\xi_1;{\bf r}_2,\xi_2)\, \frac{1}{2}\,
    \sigma({\bf r}_2)\,
    {\cal K}_0({\bf r}_2,\xi_2;{\bf \bar r},\bar{y}_l)\, .
 \label{4.35} 
\end{eqnarray}
The $N$-th order in opacity is obtained by expanding the
integrand of the gluon energy distribution (\ref{4.1})
to $N$-th order in $\left( n(\xi)\, \sigma({\bf r})\right)^N$
\cite{Wiedemann:2000za,Gyulassy:2000er,Gyulassy:2000fs}.
We start with a discussion of the first order term for 
which the medium acts effectively as a single hard momentum
transfer positioned within path length $L$. 

%
\subsubsection{Expansion to order $N=0$ and $N=1$}
For the gluon energy distribution (\ref{4.1}) of a free incoming
quark, the zeroth order in opacity vanishes. A free incoming particle 
does not radiate without interaction. A parton produced in some 
hard process, however, does radiate without further interaction
in order to decrease its virtuality. This is the DGLAP parton
shower whose medium modification is accessed by the medium-induced
spectrum $\omega \frac{dI}{d\omega}$. 
To zeroth order in $L\, n_0$, this spectrum is given by
\begin{eqnarray}
  &&\omega {d^3I(N=0)\over d\omega\, d{\bf k}}
  = {\alpha_s\over \pi^2}\, 
    C_R\, \frac{1}{{\bf k}^2}\, ,
    \qquad   H({\bf k}) =  \frac{1}{{\bf k}^2}\, ,
  \label{4.36}\\
%
  && \hspace{-1cm}
   \epsfxsize=5.0cm 
    \centerline{\epsfbox{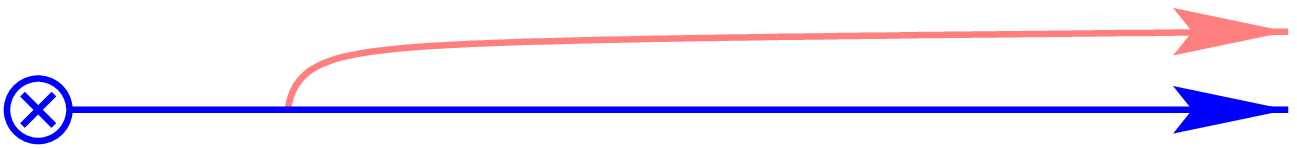}}
    \label{4.37}
\end{eqnarray} 
This is the characteristic gluon radiation spectrum associated to 
the production of a hard parton. Medium modifications to (\ref{4.36})
appear to first order in opacity.
Introducing the transverse energies $Q = \frac{{\bf k}^2}{2\omega}$, 
$ Q_1 = \frac{\left( {\bf k}  + {\bf q}_{1}\right)^2}{2\omega}$,
one finds
\begin{eqnarray}
  &&\omega {d^3I(N=1)\over d\omega\, d{\bf k}}
  = {\alpha_s\over \pi^2}\, 
    \frac{C_R}{2\,\omega^2}\, 
    \int \frac{d{\bf q}_1}{(2\pi)^2}\,
    |a_0({\bf q}_1)|^2
   \nonumber \\
    && \qquad \qquad \qquad \times
    \frac{-{\bf k}\cdot {\bf q}_1}{Q\, Q_1}\,
    n_0\, \frac{LQ_1 - \sin(LQ_1)}{Q_1}\, .
  \label{4.38}\\
   &&\epsfxsize=8.0cm 
    \centerline{\epsfbox{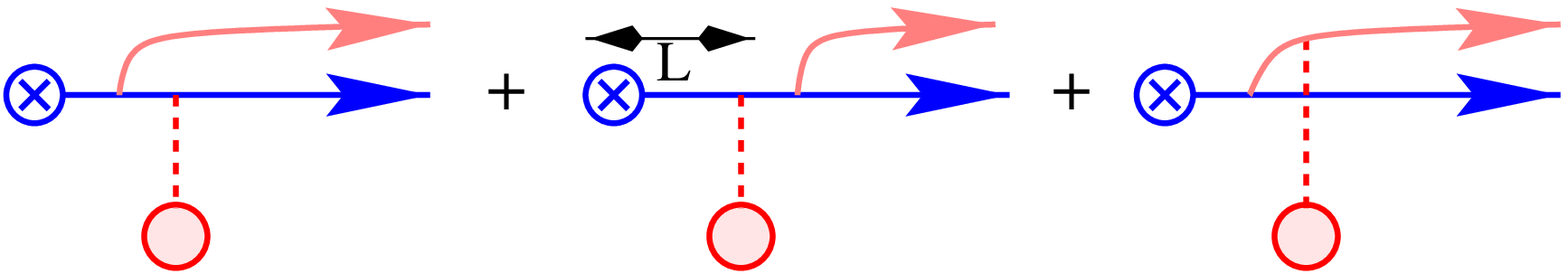}}
    \label{4.39}
\end{eqnarray} 
The formation time for a medium-induced gluon is
$t_f \approx 1/Q_1$. For a finite distance $\simeq L$ between the
production point of the parton and the point of its rescattering,
there are thus gluons whose formation times are too long: 
$L/t_f = LQ_1 < 1$. The interference term 
$\frac{LQ_1 - \sin(LQ_1)}{Q_1}$ in (\ref{4.38}) prevents 
such gluons from being radiated. It provides a logarithmic
cut-off for an otherwise infrared divergent ${\bf k}$-
integrated expression, see eq. (\ref{4.49}) below.

This radiation spectrum interpolates between the totally
coherent and totally incoherent limiting cases. These can 
be discussed in terms of the isolated contributions for the 
hard (vacuum) and the medium-induced Gunion-Bertsch radiation
terms:
\begin{eqnarray}
   &&\epsfxsize=8.0cm 
    \centerline{\epsfbox{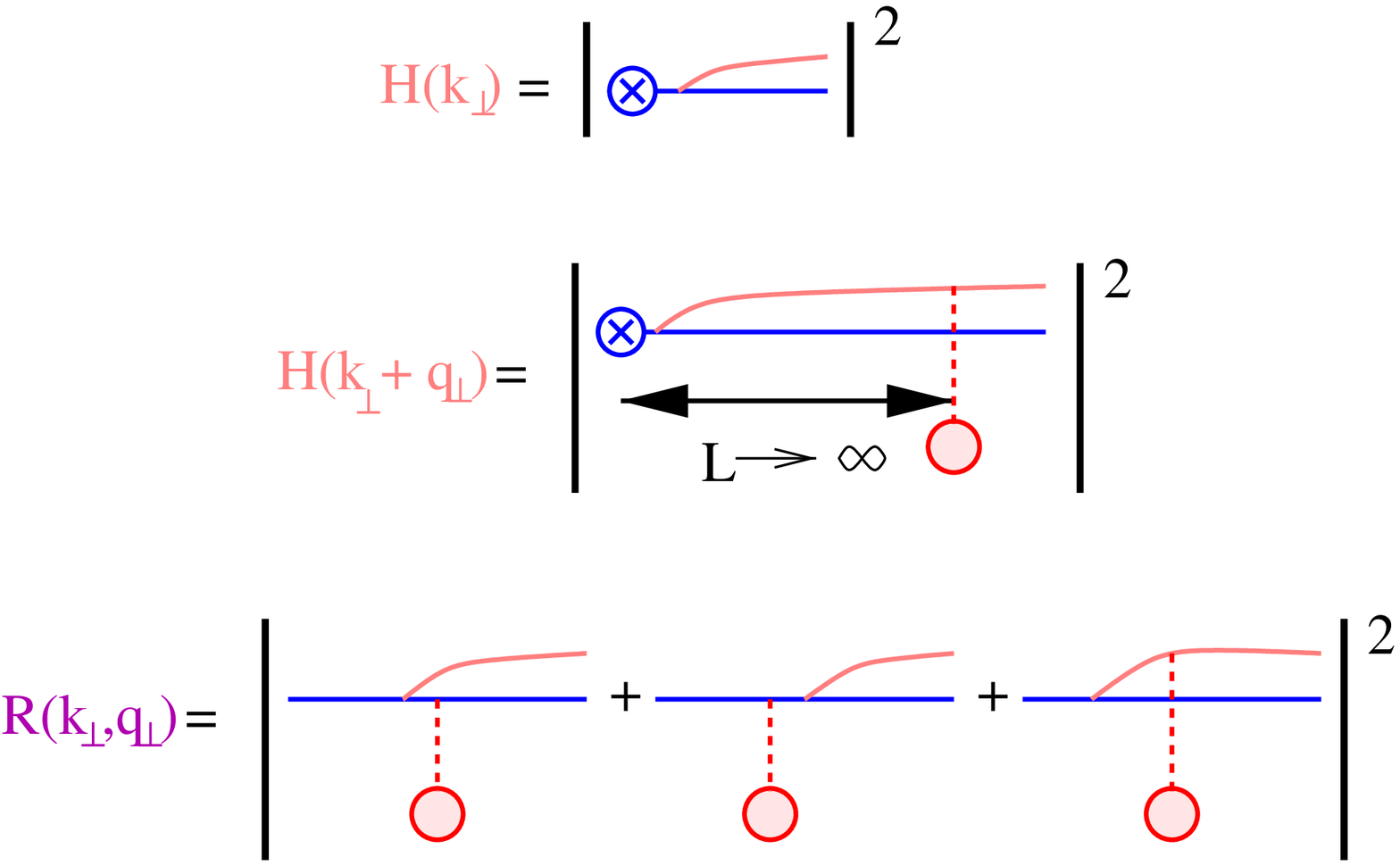}}
    \label{4.40}
\end{eqnarray} 
In the totally coherent limit, the extension of the target
is negligible, and medium-corrections are absent,
\begin{equation}
  \lim_{L\to 0}{d^3\sigma^{(nas)}(N=1)\over d(\ln x)\, d{\bf k}}
  = 0\, .  
  \label{4.41}
\end{equation}
In the incoherent $L\to\infty$ limit, the radiation cross section
(\ref{4.1}) expanded up to first order in opacity, takes the form
\begin{eqnarray}
  &&\lim_{L\to\infty}\sum_{m=0}^{N=1}
  \omega {d^3I(m)\over d\omega\, d{\bf k}}
  = {\alpha_s\over \pi^2}\, C_F
  (1-w_1)H({\bf k}) 
  \nonumber \\
  &&\qquad + {\alpha_s\over \pi^2}\, C_F
                  \, n_0\, L\, \int
		  \frac{d{\bf q}_1}{(2\pi)^2}
		  \vert a_0({\bf q}_1)\vert^2
		  \Bigg (
                  H({\bf k} + {\bf q}_{1})
                  + R({\bf k},{\bf q}_{1}) \Bigg )\, .
  \label{4.42}
\end{eqnarray}
The three terms on the r.h.s. of (\ref{4.42}) have a simple physical 
meaning: the first is the hard, medium-independent radiation 
(\ref{4.36}) reduced by the probability $w_1$ that
one interaction of the projectile occurs in the medium. The
second term describes the hard radiation component which rescatters
once in the medium. The third term is the medium-induced Gunion-Bertsch
contribution associated with the rescattering.  For the case $N=1$, 
the above discussion that eq. (\ref{4.38})
pattern interpolates between simple and physically intuitive limiting 
cases. In general, the $N$-th order term in the opacity expansion 
is a convolution of the radiation associated to $N$-fold scattering and 
a readjustment of the probabilities that rescattering occurs with less
than $N$ scattering centers~\cite{Wiedemann:2000za}.

\subsubsection{Qualitative estimates of $\omega \frac{dI}{d\omega}$
vs. quantitative calculations}
\label{sec4b2}
In general, the radiation spectrum $\omega \frac{dI}{d\omega}$
displays an interference pattern between medium-induced
and hard (vacuum) radiation amplitudes. Formation time
effects determine this interplay and allow for some
qualitative estimates. In the simplest case, consider a hard 
partonic projectile which picks up a single transverse momentum 
$\mu$ by interacting with a 
single hard scatterer. An additional gluon of energy $\omega$ decoheres
from the projectile wave function if its typical formation
time $\bar{t}_{\rm coh} = \frac{2\omega}{\mu^2}$ is smaller than the
typical distance $L$ between the production point of the parton
and the position of the scatterer. The relevant phase is 
\begin{equation}
  \gamma = \frac{L}{\bar{t}_{\rm coh}} \equiv \frac{\bar{\omega}_c}{\omega}
  \, .
  \label{4.43}
\end{equation}
Thus one expects that the radiation of gluons is suppressed
if their energy $\omega$ is larger than the characteristic gluon energy 
\begin{equation}
  \bar\omega_c = \frac{1}{2} \mu^2\, L\, .
  \label{4.44}
\end{equation}
From this one can estimate the gluon energy spectrum per unit path length 
in terms of the coherence time $\bar{t}_{\rm coh}$,
\begin{equation}
  \omega \frac{dI^{N=1}}{d\omega\, dz} \simeq 
  \frac{\alpha_s}{\bar{t}_{\rm coh}}
  \simeq \alpha_s\, \frac{\mu^2}{\omega}\, .
  \label{4.45}
\end{equation}
This estimate can be compared to the full $N=1$ opacity
term (\ref{4.38}) integrated over transverse momentum up
to $\vert {\bf k}\vert < \chi\, \omega$. Introducing the
kinematic constraint in transverse momentum phase space,
\begin{equation}
  \bar{R} = \chi^2 \bar\omega_c\, L\, ,
  \label{4.46}
\end{equation}
the gluon radiation spectrum takes the 
form\cite{Salgado:2003gb}
\begin{eqnarray}
  \omega \frac{dI^{N=1}}{d\omega} &=& 2 \frac{\alpha_s\, C_R}{\pi}\,
   (n_0L)\, \gamma\,   
  \int_0^\infty  dr\,  \frac{r - sin(r)}{r^2}
                 \nonumber \\   
  && \times
  \left( \frac{1}{r + \gamma} - 
         \frac{1}{\sqrt{( (\bar{R}/2\gamma) + r + \gamma)^2
                       - 4 r\bar{R}/2\gamma}}\right)\, .
  \label{4.47}
\end{eqnarray}
%
%
\begin{figure}[t]\epsfxsize=10.7cm
\centerline{\epsfbox{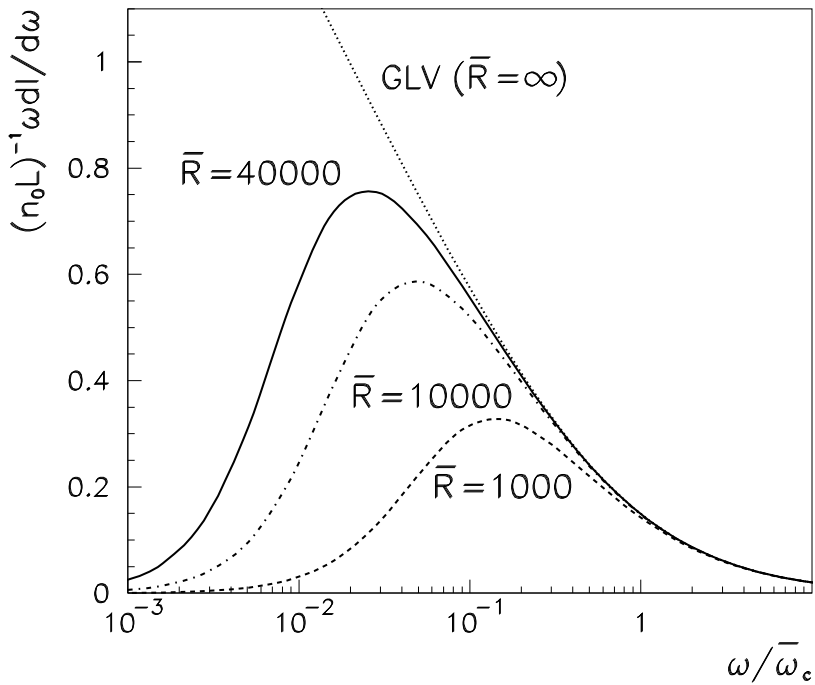}}
\caption{
The medium-induced gluon energy distribution 
$\omega \frac{dI}{d\omega}$ in the $N=1$ opacity 
expansion. Different curves are for different values of the
kinematic constraint $\bar{R} = \bar{\omega}_c\, L$.
Figure taken from \protect\cite{Salgado:2003gb}. 
}\label{fig6}
\end{figure}
%
The expression (\ref{4.47}) is plotted in Fig.~\ref{fig6}. It shows 
qualitatively the
same shape as the spectrum in the multiple soft scattering
case. The position of the maximum of $\omega \frac{dI(N=1)}{d\omega}$
changes $ \propto \frac{1}{\sqrt{\bar{R}}}$. This is in
accordance with the assumption that gluon radiation is suppressed 
for energies below which the characteristic angle of the
gluon emission is of order one
\begin{equation}
  \Theta_c^2 \simeq \frac{\mu^2}{\omega^2}
  \simeq \left(\frac{\bar\omega_c}{\omega} \right)^2 
         \frac{1}{\bar{R}} \sim 1\quad \Longrightarrow
         \quad \frac{\omega}{\bar\omega_c} 
               \propto \frac{1}{\sqrt{\bar{R}}}  \, .
  \label{4.48}
\end{equation}

In the limit $\bar{R} \to \infty$ in which the phase
space constraint is removed, one recovers some pocket formulas
from the literature~\cite{Gyulassy:2000fs}. First, this limit 
satisfies the
characteristic $1/\omega$-energy dependence of the 
estimate (\ref{4.45}) for sufficiently large gluon energies 
$\omega > \bar\omega_c$,
\begin{eqnarray}
   \lim_{\bar{R}\to \infty}\, 
   \omega \frac{dI(N=1)}{d\omega} &=& 
   2\, \frac{\alpha_s\, C_R}{\pi}\, \left( n_0\, L\right)\,
   \gamma\,   
  \int_0^\infty  dr\, \frac{1}{r + \gamma}\,  
                  \frac{r - sin(r)}{r^2}\, 
   \nonumber \\
   &\simeq & 
   2\, \frac{\alpha_s\, C_R}{\pi}\, \left( n_0\, L\right)\,
          \left\{ \begin{array} 
                  {r@{\qquad  \hbox{for}\quad}l}
                  \log \left[ \frac{\bar\omega_c}{\omega}\right]
                  & \bar\omega_c > \omega\\ 
                  \frac{\pi}{4}\, \frac{\bar\omega_c}{\omega}
                  & \bar\omega_c < \omega  
                  \end{array} \right.
  \label{4.49}
\end{eqnarray}
Moreover, this expression allows to estimate the average parton 
energy loss for a single hard scattering. Remarkably, this seems
dominated by contributions from the region 
$\omega > \bar\omega_c$,\cite{Gyulassy:2000fs,Zakharov:2000iz}
\begin{equation}
 \lim_{\bar{R} \to \infty} \langle \Delta E \rangle^{N=1} = 
 \lim_{\bar{R} \to \infty}\,  \int d\omega\, 
   \omega \frac{dI^{N=1}}{d\omega}
   \simeq \frac{\alpha_s C_R}{2} (n_0L)\, \bar\omega_c\, 
   \log\left[ E/\bar\omega_c\right]\, , 
 \label{4.50}
\end{equation}
which appears to be logarithmically enhanced in comparison
to the region $\omega < \bar{\omega}_c$ for which
\begin{equation}
 \lim_{\bar{R} \to \infty}\,  \int_0^{\bar{\omega}_c} d\omega\, 
   \omega \frac{dI^{N=1}}{d\omega}
   \simeq \frac{2\, \alpha_s C_R}{\pi} (n_0L)\, \bar\omega_c\, . 
 \label{4.51}
\end{equation}
However, the logarithmic enhancement of (\ref{4.50}) does not
persist upon closer inspection: All calculations so far worked
for fixed coupling constant which may be justified if all
momentum transfers are of the same order. In calculating (\ref{4.50}),
however, the ${\bf k}$-integration was extended up to the total
energy $E$. If one takes the running of the coupling into account,
this logarithmic enhancement factor is reduced to a 
$\log\log$ factor.

\subsubsection{Parameters in the opacity expansion}
\label{sec4b3}

In the multiple scattering approximation, the gluon energy distribution
(\ref{4.1}) depends on two quantities: the transport coefficient
$\hat{q}$ and the in-medium path length $L$ (we ignore the angular
dependence in the following discussion). In the opacity expansion,
there are three parameters instead: the in-medium path length $L$,
the transverse momentum squared $\mu^2$ transferred on average by a 
single scattering, and the inverse mean free path $n_0 = \frac{1}{\lambda}$.
The multiple scattering approximation describes the average momentum
transfer per unit path length while the opacity expansion specifies
in addition by how many scattering centers this momentum is
transferred on average. In the limit of multiple soft scattering,
this additional information is redundant since
\begin{equation}
  \mu^2\, n_0\, L = \hat{q}\, L\, ,
  \qquad \hbox{for Brownian motion.}
  \label{4.52}
\end{equation}
However, deviations from Brownian motion arise due to the high 
transverse momentum tails of the elastic scattering cross sections
$  \vert a({\bf q})\vert^2 = 
  \left(\mu^2 (2\pi)^2\right)/
  \left({\pi ({\bf q}^2 + \mu^2)^2}\right)$.
These tails lead to a logarithmically enhancement factor in
the transport coefficient
\begin{equation}
  \hat{q}\, L = n_0 L \int^Q \frac{d^2{\bf q}}{(2\pi)^2}\, 
  \vert a({\bf q})\vert^2\, \frac{1}{2}\, {\bf q}^2\, 
  \cos^2\varphi \sim (n_0 L)\, \mu^2\, \ln
  \sqrt{\frac{Q}{\mu}}\, .
  \label{4.53}
\end{equation}
This allows to relate via 
$\omega_c \simeq (n_0L)\, \bar{\omega_c} \ln \sqrt{\frac{Q}{\mu}}$
the characteristic gluon energies (\ref{4.17}) and (\ref{4.44}) which 
set the energy scale in the multiple soft and single hard scattering 
limits. For realistic values [$\mu \geq \Lambda_{\rm QCD}$ and
$Q \leq E$ say],  $\ln \sqrt{\frac{Q}{\mu}} \ll 10$, one concludes 
from Figs.~\ref{fig1} and ~\ref{fig6} that the medium-induced gluon 
energy distribution is significantly harder in the opacity 
approximation than in the multiple soft scattering 
limit\cite{Salgado:2003gb}.

\section{Applications}
\label{sec5}
Irrespective of the number of additionally radiated gluons, 
what matters for the medium modification of hadronic observables
is how much {\it additional} energy $\Delta E$ is radiated off a 
hard parton. In this section, we first discuss the so called
quenching weight which is the probability distribution
$P(\Delta E)$ of the additional medium-induced
energy loss. For independent gluon emission, this probability
is the normalized sum of the emission probabilities
for an arbitrary number of $n$ gluons which carry away a total
energy $\Delta E$:\cite{Baier:2001yt}
\begin{eqnarray}
  P(\Delta E) = \sum_{n=0}^\infty \frac{1}{n!}
  \left[ \prod_{i=1}^n \int d\omega_i \frac{dI(\omega_i)}{d\omega}
    \right]
    \delta\left(\Delta E - \sum_{i=1}^n \omega_i\right)
    e^{- \int d\omega \frac{dI}{d\omega}}\, .
   \label{5.1}
\end{eqnarray}
Then we discuss how this probability can be used to calculate
the medium modification of hadronic observables. 

\subsection{Properties of  Quenching Weights}
\label{sec5a}

In general, the quenching weight (\ref{5.1}) has a discrete and a 
continuous part,\cite{Salgado:2002cd}
\begin{equation}
  P(\Delta E) = p_0\, \delta(\Delta E) + p(\Delta E)\, .
   \label{5.2}
\end{equation}
The discrete weight $p_0$ emerges as a consequence of a finite
mean free path. It determines the probability that 
no additional gluon is emitted due to in-medium scattering 
and hence no medium-induced energy loss occurs. 

In order to determine the discrete and continuous part of
(\ref{5.2}), it is convenient to rewrite eq. (\ref{5.1}) 
as a Laplace transformation~\cite{Baier:2001yt}
\begin{eqnarray}
  P(\Delta E) &=& \int_C \frac{d\nu}{2\pi i}\, {\cal P}(\nu)\,
  e^{\nu\Delta E}\, ,
  \label{5.3}\\
  {\cal P}(\nu) &=& \exp\left[ -\int_0^\infty
    d\omega\, \frac{dI(\omega)}{d\omega}\,
    \left(1- e^{-\nu\, \omega}\right)\right]\, .
  \label{5.4}
\end{eqnarray}
Here, the contour $C$ runs along the imaginary axis with
${\rm Re}\nu = 0$. 

For the further discussion, it is useful to treat the
medium-induced gluon energy distribution $\omega \frac{dI}{d\omega}$ 
in eq. (\ref{4.1}) explicitly as the medium modification of a
``vacuum'' distribution \cite{Salgado:2003gb}
\begin{equation}
 \omega \frac{dI^{({\rm tot})}}{d\omega} = 
 \omega \frac{dI^{({\rm vac})}}{d\omega} +
 \omega \frac{dI}{d\omega}\, .
 \label{5.5}
\end{equation}
From the Laplace transform (\ref{5.3}), one finds the 
total probability
\begin{equation}
  P^{({\rm tot})}(\Delta E) = \int_0^\infty  d\bar{E} \, 
  P(\Delta E - \bar{E}) \, 
   P^{({\rm vac})}(\bar{E})\, . 
   \label{5.6}
\end{equation} 
This probability $P^{({\rm tot})}(\Delta E)$ is normalized to unity and it is 
positive definite. In contrast, the medium-induced modification of this 
probability, $P(\Delta E)$, is a generalized probability. It can take
negative values for some range in $\Delta E$, as long as its normalization
is unity,
\begin{equation}
  \int_0^\infty  d\bar{E} \,  P(\bar{E})
  = p_0 + \int_0^\infty  d\bar{E} \,  p(\bar{E}) = 1\, .
  \label{5.7}
\end{equation}
We now discuss separately the properties of the discrete contribution
$p_0$ and the continuous one $p(\bar{E})$. A CPU-inexpensive Fortran
routine\cite{Salgado:2003gb} is available for the calculation of these 
quenching weights.

\begin{figure}[h]\epsfxsize=9.7cm
\centerline{\epsfbox{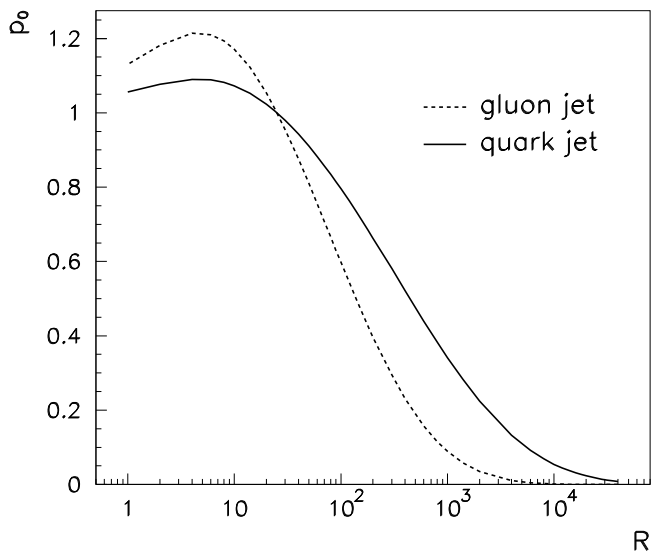}}
\caption{The discrete part $p_0$ of the quenching weight calculated
in the multiple soft scattering limit as a function of $R$.
Figure taken from \protect\cite{Salgado:2003gb}. 
}\label{fig7}
\end{figure}
\subsubsection{Discrete part of the quenching weight}
The discrete part of the quenching weight is the $n=0$ term of
eq. (\ref{5.1}). It 
can be expressed in terms of the total gluon multiplicity,
\begin{equation}
  p_0 = \lim_{\nu \to \infty}\, {\cal P}(\nu) = 
  \exp \left[ -N(\omega = 0)\right]\, ,
  \label{5.8}
\end{equation}
where the multiplicity $N(\omega)$ of gluons with energy larger 
than $\omega$ emerges by partially integrating the exponent of
(\ref{5.4}),
\begin{equation}
  N(\omega) \equiv \int_\omega^\infty d\omega'\,
                    \frac{dI(\omega')}{d\omega'}\, .
  \label{5.9}
\end{equation}
For the limiting case of infinite in-medium path length, the
total multiplicity $N(\omega)$ diverges and the discrete 
part vanishes. In general, however, $p_0$ is finite. 
A typical dependence of $p_0$ on model parameters is
shown in Fig.~\ref{fig7} for the radiation spectrum
calculated in the multiple soft scattering limit.
A qualitatively similar behavior is found in the 
opacity expansion. 
Remarkably, $p_0$ can exceed unity for some parameter
range, since the medium modification $\omega \frac{dI}{d\omega}$
to the radiation spectrum (\ref{5.5}) can be negative.  
The value $p_0> 1$ then compensates a predominantly 
negative continuous part $p(\Delta E)$ and satisfies 
the normalization (\ref{5.7}). This indicates
a phase space region at very small transverse momentum, 
into which {\it less} gluons 
are emitted in the medium than in the vacuum. 
This effect is more pronounced for gluons than for quarks. 

\subsubsection{Continuous part of the quenching weight}
\label{sec5a2}
The continuous part $p(\Delta E)$ of the probability distribution 
(\ref{5.2}) can be calculated numerically from the Mellin transform
(\ref{5.4}). To facilitate the numerical calculation, one subtracts 
from $P(\nu)$ the discrete contribution $p_0$ which dominates the 
large-$\nu$ behavior.

\begin{figure}[t]\epsfxsize=14.0cm
\centerline{\epsfbox{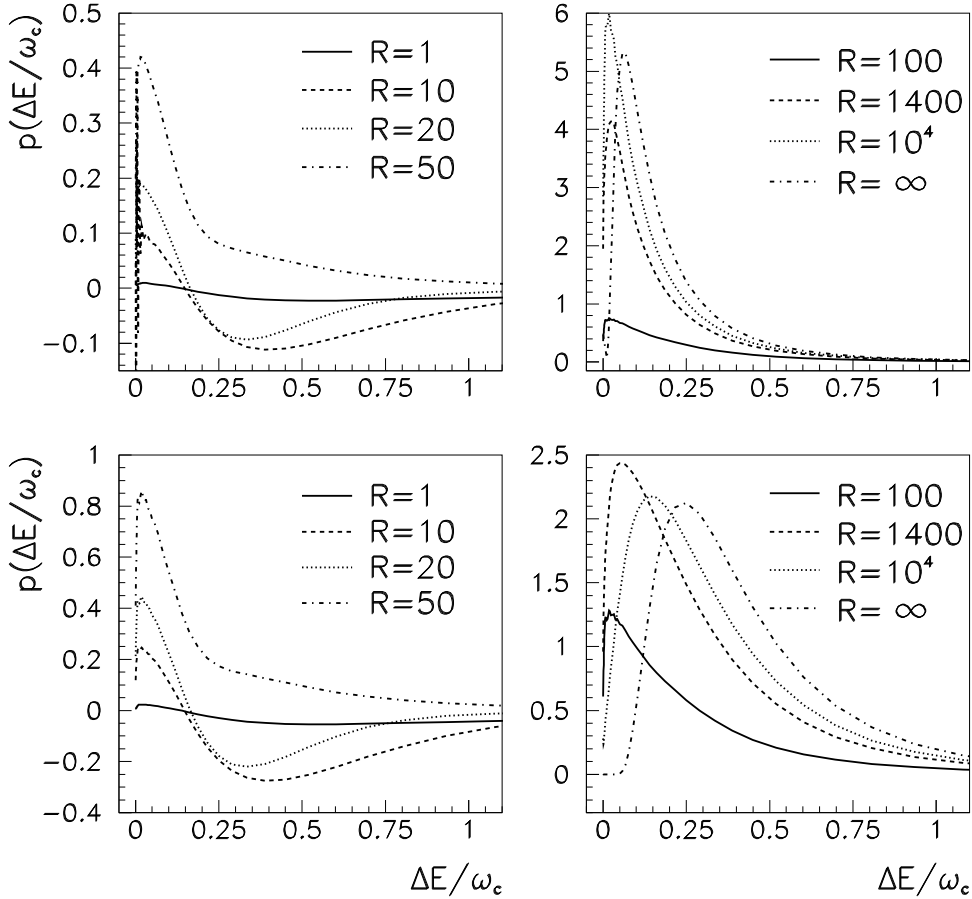}}
\caption{The continuous part of the quenching weight 
(\protect\ref{5.2}), calculated in the multiple soft scattering limit
for a hard quark (upper row) or hard gluon (lower row).
Figure taken from \protect\cite{Salgado:2003gb}.
}\label{fig8}
\end{figure}

Fig.~\ref{fig8} shows the continuous part of the quenching weight,
calculated in the multiple soft scattering limit. In the opacity expansion,
it looks qualitatively similar. As expected from the normalization
condition (\ref{5.7}), the continuous part $p(\Delta E)$ shows
predominantly negative contributions for the parameter range for
which the discrete weight $p_0$ exceeds unity. 
With increasing density of
scattering centers (i.e. increasing $R = \frac{1}{2}\hat{q}L^3$)
the probability of loosing a significant energy fraction
$\Delta E$ increases. The energy loss is larger for gluons 
which have a stronger coupling to the medium. This broadens
the width of $p(\Delta E)$ for the gluonic case. 
%

In the multiple soft scattering approximation, an analytic estimate 
for the quenching weight can be obtained\cite{Baier:2001yt} in the 
limit $R\to \infty$ from the small-$\omega$ approximation 
$\omega \frac{dI}{d\omega} \propto \frac{1}{\sqrt{\omega}}$ 
\begin{equation}
  P^{\rm approx}_{\rm BDMS}(\epsilon)
  = \sqrt{\frac{a}{\epsilon^3}} 
     \exp\left[-\frac{\pi\, a}{\epsilon} \right]\, ,
     \qquad \hbox{\rm where}\, \, 
     a=\frac{2\, \alpha_s^2\, C_R^2}{\pi^2}\omega_c\, .
  \label{5.10}
\end{equation}
This reproduces roughly \cite{Salgado:2002cd} the 
shape of the probability distribution for large system size, but 
it has an unphysical large $\epsilon$-tail with infinite first moment 
$\int d\epsilon\, \epsilon\, P^{\rm approx}_{\rm BDMS}(\epsilon)$.
An alternative analytic approach \cite{Arleo:2002kh} aims at fitting 
a two-parameter log-normal distribution to the numerical result
for $P(\Delta E)$. 

\subsection{Quenching factors for hadronic spectra}
\label{sec5b}
Assume that a hard parton looses an additional energy fraction
$\Delta E$ while escaping the collision region. 
The medium-dependence of the corresponding inclusive transverse momentum
spectra can be characterized in terms of the quenching 
factor $Q$ \cite{Baier:2001yt}
\begin{eqnarray}
 Q(p_\perp)&=&
{{d\sigma^{\rm med}(p_\perp)/ dp^2_\perp}\over
{d\sigma^{\rm vac}(p_\perp)/ dp^2_\perp}}=
\int d{\Delta E}\, P(\Delta E)\left(
{d\sigma^{\rm vac}(p_\perp+\Delta E)/ dp^2_\perp}\over
{d\sigma^{\rm vac}(p_\perp)/ dp^2_\perp}\right)
\nonumber\\[2mm]
&\simeq&  \int d{\Delta E}\, P(\Delta E)\,  
    \left({p_\perp\over p_\perp+\Delta E}\right)^n\, .
 \label{5.11}
\end{eqnarray}
Here, the last line is obtained by assuming a power law fall-off of 
the $p_\perp$-spectrum. The effective power $n$ depends in general
on $p_\perp$. It is $n \simeq 7$ for the $p_\perp$-range relevant
for RHIC. Alternatively, instead of the quenching factor (\ref{5.11}),
the medium modification of hadronic transverse momentum spectra is often 
characterized by a shift factor $S(p_\perp)$,
\begin{equation}
  \frac{d\sigma^{{\rm med}}(p_\perp)}{dp^2_\perp}  \simeq
  \frac{d\sigma^{{\rm vac}}(p_\perp+S(p_\perp))}{dp^2_\perp} \, , 
  \label{5.12}
\end{equation} 
which is related to the shift $S(p_\perp)$ by
\begin{equation}
  Q(p_\perp) = \exp\left\{-\frac{n}{p_\perp}\cdot
  S(p_\perp)\right\} \, .
  \label{5.13}
\end{equation}
Most importantly, since the hadronic spectrum shows a strong
power law decrease, what matters for the suppression is not
the average energy loss $\langle \Delta E \rangle$ but 
the least energy loss with which a hard parton is likely
to get away. One concludes that $S(p_\perp) < \langle \Delta E \rangle$
and depends on transverse momentum~\cite{Baier:2001yt}. 
%
\begin{figure}[h]\epsfxsize=14.7cm
\centerline{\epsfbox{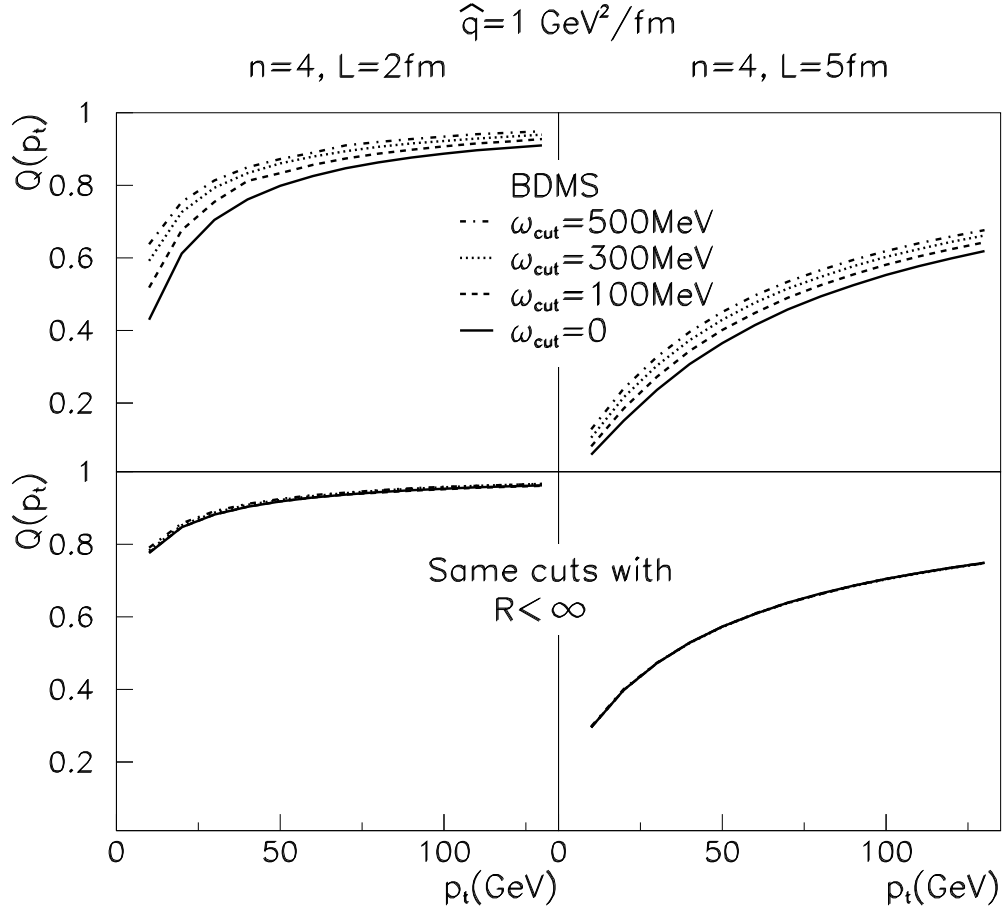}}
\caption{The quenching factor (\protect\ref{5.11}) calculated in the
multiple soft scattering limit. Upper row: calculation in the 
$R\to\infty$-limit but with a variable sharp cut-off on the 
infrared part of the gluon energy distribution. Lower row: the 
same calculation is insensitive to infrared contributions if 
the finite kinematic constraint $R=\omega_cL <\infty$ is included. 
Figure taken from \protect\cite{Salgado:2003gb}.
}\label{fig9}
\end{figure}
%
Fig.~\ref{fig9} shows a calculation of the quenching factor
(\ref{5.11}) in the multiple soft scattering limit. A 
qualitatively similar result is obtained in the opacity
expansion. In general, quenching weights increase
monotonically with $p_\perp$ since the medium-induced
gluon radiation is independent of the total projectile energy
for sufficiently high energies. At very low transverse momenta,
the calculation based on (\ref{4.1}) is not reliable
and the interpretation of the medium modification of
hadronic spectra in nucleus-nucleus collisions will 
require additional input (e.g. modifications due to the
Cronin effect). Fig.~\ref{fig9} suggests, however, that
hadronic spectra at transverse momenta $p_\perp > 10$ GeV,
can be suppressed significantly due to partonic final state
rescattering. 

To quantify the sensitivity of the calculation to the low momentum 
region, Baier et al.\cite{Baier:2001yt} introduced a sharp 
cut-off on the $R\to \infty$ gluon energy distribution which was
varied between $\omega_{\rm cut} = 0$ and $\omega_{\rm cut} = 500$ MeV.
However, phase space constraints (i.e. finite $R$) deplete the
gluon radiation spectrum in the soft region, see Fig.~\ref{fig2}.
As seen in Fig.~\ref{fig9}, this decreases significantly the 
sensitivity of quenching factors to the uncontrolled infrared
properties of the radiation spectrum. 

\subsection{Medium-modified fragmentation functions}
\label{sec5c}
For an alternative calculation of the medium modification of
hadronic spectra, one may determine the dependence of fragmentation 
functions on partonic energy loss. In general, hadronic
cross sections are calculated by convoluting the parton
distributions of the incoming projectiles with the product $d\sigma^h(z,Q^2)$
of a perturbatively calculable partonic cross section $\sigma^q$ and
the fragmentation function $D_{h/q}(x,Q^2)$ of the produced parton,
$d\sigma^h(z,Q^2) = \left( \frac{d\sigma^q}{dy}\right) dy\,
D_{h/q}(x,Q^2)\, dx$. Here, $x=E_h/E_q$, $y=E_q/Q$ and $z=E_h/Q$
denote fractions between the virtuality of the hard process $Q$,
and the energies of the produced parton and resulting hadron.
If the produced parton loses with probability $P(\epsilon)$
an additional fraction $\epsilon = \frac{\Delta E}{E_q}$  of
its energy due to medium-induced radiation, then the hadronic
cross section is given in terms of the medium-modified
fragmentation function~\cite{Wang:1996yh,Gyulassy:2001nm}
\begin{eqnarray}
  D_{h/q}^{(\rm med)}(x,Q^2) = \int_0^1 d\epsilon\, P(\epsilon)\,
  \frac{1}{1-\epsilon}\, D_{h/q}(\frac{x}{1-\epsilon},Q^2)\, .
  \label{5.14}
\end{eqnarray}
The hadronized remnants
of the medium-induced soft radiation are neglected in the definition
of (\ref{5.14}). However, these remnants are expected to be
soft, and their inclusion would thus amount to an additional 
contribution to $D_{h/q}^{(\rm med)}(x,Q^2)$ for
$x > 0.1$ say. 
%
\begin{figure}[h]\epsfxsize=8.7cm
\vspace{-1.0cm}
\centerline{\epsfbox{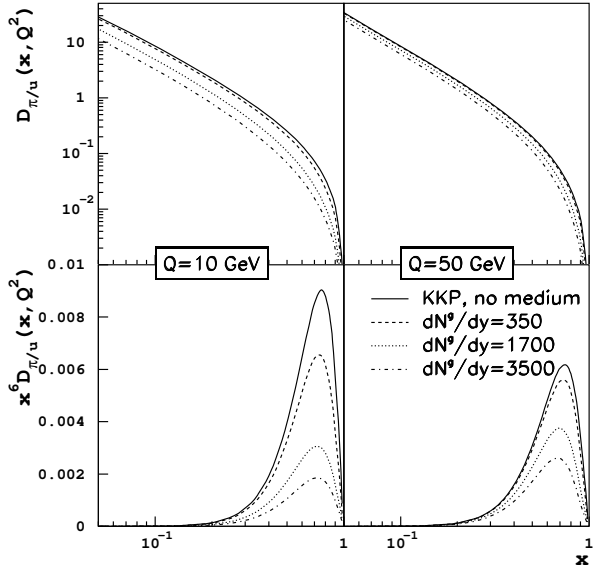}}
\caption{The LO KKP~\protect\cite{Kniehl:2000fe} fragmentation function
 $u\to \pi$ for no medium and the medium-modified fragmentation
functions for different gluon rapidity densities 
(see eq. (\protect\ref{5.15}))
and $L=7$ fm. Figure taken from \protect\cite{Salgado:2002cd}.
}\label{fig10}
\end{figure}

Fig.~\ref{fig10} shows a calculation of the parton fragmentation
functions $D_{\pi/q}(x,Q^2)$ from (\ref{5.14}) using the quenching
weights of Fig.~\ref{fig8} and the LO KKP~\cite{Kniehl:2000fe}
parametrization of $D_{h/q}(x,Q^2)$. For this calculation, 
the virtuality $Q$ of $D_{h/q}(x,Q^2)$ is identified with
the (transverse) initial energy $E_q$ of the parton.
This is justified since $E_q$ and $Q$ are of the same order, and
$D_{h/q}(x,Q^2)$ has a weak logarithmic $Q$-dependence while
medium-induced effects change as a function of
$\epsilon = \frac{\Delta E}{Q} \approx O(\frac{1}{Q})$.
For a collision region expanding according to Bjorken scaling, 
the transport coefficient can be related to the initial gluon 
rapidity density~\cite{Baier:1996sk,Gyulassy:2000gk},  
\begin{equation}
  R = \frac{1}{2}\hat{q}L^3 = \frac{L^2}{R_A^2}\, \frac{dN^g}{dy}\, .
\label{5.15}
\end{equation}
That's what is done in Fig.~\ref{fig10}. Interestingly, eq. (\ref{5.15})
indicates how partonic energy loss changes with the particle 
multiplicity in nucleus-nucleus collisions. This allows to 
extrapolate parton energy loss effects from RHIC to LHC
energies~\cite{Salgado:2002cd}.

In principle, the medium modified fragmentation function
should be convoluted with the hard partonic cross section
and parton distribution functions in order to determine the
medium modified hadronic spectrum. For illustration, however,
one may exploit that hadronic cross sections 
weigh $D_{h/q}^{(\rm med)}(x,Q^2)$ by the partonic cross section
${d\sigma^q}/{dp_{\perp}^2} \sim {1}/{p_{\perp}^{n(\sqrt{s}, p_{\perp})}}$
and thus effectively test $x^{n(\sqrt{s}, p_{\perp})} 
D_{h/q}^{(\rm med)}(x,Q^2)$~\cite{Eskola:2002kv}. The value
$n = 6$ characterizes~\cite{Eskola:2002kv} the power law
for typical values at RHIC ($\sqrt{s} = 200$ GeV and $p_\perp \sim 10$ GeV).
Thus, the
position of the maximum $x_{\rm max}$ of  $x^6 D_{h/q}^{(\rm med)}(x,Q^2)$
corresponds to the most likely energy fraction $x_{\rm max} E_q$
of the leading hadron. And the suppression around its maximum translates 
into a corresponding relative suppression of this contribution to the
high-$p_\perp$ hadronic spectrum at $p_\perp \sim x_{\rm max} E_q$.
In general, the suppression of hadronic spectra extracted in this
way is in rough agreement\cite{Salgado:2003gb} with calculations of 
the quenching factor (\ref{5.11}).

\section{Appendix A: Eikonal calculations in the target light cone gauge.}
In section~\ref{sec2} of this review we have used the light cone gauge 
$A^-=0$. In this gauge the gluon and quark distributions of the projectile 
wave function are simply expressible in terms of the gluon and quark 
number operators and we will therefore refer to it as the projectile 
light cone gauge (PLCG). The standard light cone gauge used in DIS 
calculations on the other hand is $A^+=0$, which facilitates simple 
expressions of the target distribution functions. In this appendix we 
show how the spectrum of emitted gluons is obtained in this standard 
light cone gauge, which we will refer to as the target light cone gauge 
(TLCG).

Recall, that in PLCG the target is described as an ensemble of gluon 
fields with dominant component $A^+$. The eikonal $S$-matrix for the 
propagation of a charged parton through the target is given my the 
Wilson line eq.(\ref{2.3})
\begin{equation}
  W({\bf x}_i)={\cal P}\exp\{i\int dz_-T^aA^+_a({\bf x}_i, z_-)\}
\label{a.1}
\end{equation}
In the TLCG the $A^+$ component of the vector potential vanishes. 
Instead the chromoelectric fields in the target are given in terms 
of the transverse components $A_i$. The $A^i$ are obtained from $A^+$ 
by the gauge transformation from PLCG to TLGT \cite{Kovner:2000pt}
\begin{equation}
A_i({\bf x}_i,x_-)=iV^\dagger({\bf x}_i,x_-)\partial_iV({\bf x}_i,x_-)
\label{a.2}
\end{equation}
where
\begin{equation}
  V({\bf x}_i,x_-)={\cal P}\exp\{i\int_{-\infty}^{x_-} 
  dz_-T^aA^+_a({\bf x}_i, z_-)\}
\label{a.3}
\end{equation}
Note that the TLGT condition $A^+=0$ does not fix the gauge 
unambiguously, but only up to residual gauge transformation which 
does not depend on $x^-$. The choice of the lower limit of the $z_-$ 
integration in eq.(\ref{a.3}) is equivalent to fixing this residual 
gauge freedom by imposing the condition 
$\partial_jA_j({\bf x}_i,x_-\rightarrow-\infty)=0$ \cite{Kovner:2000pt}.
With this choice we have
\begin{equation}
  V({\bf x}_i,x_-\rightarrow-\infty)=1, \ \ \ \ 
  V({\bf x}_i,x_-\rightarrow\infty)=W({\bf x}_i)\, .
  \label{a.4}
\end{equation}

Eq.(\ref{a.2}) defines the vector potential $A_i$ 
as two dimensional pure gauge, 
$\partial_iA^a_j-\partial_jA^a_i-f^{abc}A^b_iA^c_j=0$. Moreover, 
at $x_-\rightarrow +\infty$ the vector potential is genuinely (and not
just two dimensionally) pure gauge.

Let us now consider scattering of a projectile with the wave function 
eq.(\ref{2.1}) on the target described by ensemble of fields $A_i$ of the 
form eq.(\ref{a.2}).
Since the $A^+$ component of the vector potential vanishes, in eikonal 
approximation the wave function of the projectile {\it does not change}
while it propagates through the target. The outgoing wave function therefore
is equal to the incoming one
\begin{equation}
  \Psi_{out} = \sum_{\{\alpha_i,{\bf x}_i\}}\, \psi(\{\alpha_i, {\bf x}_i\})\, 
  \vert\{\alpha_i,{\bf x}_i\}\rangle\, .
\label{a.5}
\end{equation}

This however does not mean that no scattering takes place. To calculate 
the scattering amplitude one has to project the outgoing
wave function into the Hilbert space orthogonal to the wave function of 
the freely propagating system far "to the right" of the target, that is
at $x_-\rightarrow+\infty$.
In the PLCG the target gauge field vanishes at both $x_-\rightarrow-\infty$ 
and $x_-\rightarrow+\infty$ and the freely propagating wave functions 
are identical "to the left" and "to the right" of the target. However 
in TLGT this is not the case. At $x_-\rightarrow+\infty$ the target 
vector potential does not vanish, but is instead a pure gauge, 
eq. (\ref{a.2}). Therefore the freely propagating wave function
at $x_-\rightarrow+\infty$ is not identical to that at $x_-\rightarrow-\infty$,
but is rather its gauge transform with the gauge transformation generated by
the Wilson loop eq. (\ref{a.1}). 
In particular the fields "to the right" of the target, $A$ are related
to the fields "to the left" of the target, $a$ by
\begin{equation}
  A=W^\dagger aW+iW^\dagger\partial W\, .
  \label{a.6}
\end{equation}
Thus the free Fock basis "to the right" of the target 
is related to the free Fock basis "to the left" of the target by
\begin{equation}
  \vert\{\alpha,{\bf x}_i\}\rangle_R=
  W^\dagger_{\alpha\beta}({\bf x}_i)\vert\{\beta,{\bf x}_i\}\rangle_L\ .
  \label{a.7}
\end{equation}
The outgoing wave function eq.(\ref{a.5}) is given in the basis
$\vert\{\alpha,{\bf x}_i\}\rangle_L$. On the other hand all observables 
at late time after scattering, including the number of emitted
gluons must be calculated with respect to the basis 
$\vert\{\alpha,{\bf x}_i\}\rangle_R$. It is thus convenient to rewrite
$\Psi_{out}$ using eq.(\ref{a.7}) as
\begin{equation}
  \Psi_{out}=
  \sum_{\{\alpha_i,{\bf x}_i\}}\psi(\{\alpha_i, {\bf x}_i\})
  \prod_iW({\bf x}_i)_{\alpha_i \beta_i}\, 
  \vert\{\beta_i,{\bf x}_i\}\rangle_R\, .
  \label{a.8}
\end{equation}
However the same projectile freely propagating to the right of the target
would have the wave function
\begin{equation}
  \Psi_{free} = \sum_{\{\alpha_i,{\bf x}_i\}}\, 
  \psi(\{\alpha_i, {\bf x}_i\})\, 
  \vert\{\alpha_i,{\bf x}_i\}\rangle_R\, .
\label{a.9}
\end{equation}

Thus we see that the outgoing wave function indeed differs from the 
freely propagating one, and thus the scattering is nontrivial and 
gluons are emitted in the final state. It is a somewhat curious feature 
of the TLCG that the nontrivial scattering amplitude appears entirely 
due to rotation of the free particle basis between early and late times
Nevertheless it is obvious that the results of the calculation in this 
gauge are identical to those in PLCG as they should be. In fact from
this point on all calculations are identical to those presented in 
section~\ref{sec2}, as the interesting part of the outgoing wave function 
is given by eq.(\ref{2.5}) with $\Psi_{free}$ of eq.(\ref{a.9}) 
substituted for $\Psi_{in}$ and the "free" gluons are defined as 
states in the $R$ Hilbert space, $\vert\{\alpha,{\bf x}\}\rangle_R$.
        
\section{Appendix B: Path integral formalism for the photon radiation spectrum}
In section~\ref{sec3c}, we discussed how to derive the
non-abelian gluon radiation spectrum from the non-abelian
Furry approximation (\ref{3.16}). In this appendix, we
present in more detail the derivation of the abelian
analogue from the abelian Furry wave function (\ref{3.22}).
The QED photon radiation cross section in terms of Furry
wave functions reads\cite{Wiedemann:1999fq}
\begin{eqnarray}
 && \frac{d^5\sigma}{d({\rm ln}x)\,d{\bf p}\,d{\bf k}} =
 \frac{\alpha_{em}}{(2\pi)^4}\, \left|M_{fi}\right|^2\, ,
 \label{b.1}\\
 && M_{fi} = \int d^4x\,{\Psi^-}^{\dagger}(x,p_2)\,
  {\bf \alpha}\cdot {\bf \epsilon}\, e^{-\epsilon\, |z|}\,
  e^{i\, k\cdot x}\,
  \Psi^+(x,p_1)\ .
  \label{b.2}
\end{eqnarray}
Here, $e^{-\epsilon\, |z|}$ is the adiabatic switching off of the 
interaction term at large distances, discussed below eq. (\ref{3.36}).
To simplify (\ref{b.2}), we perform the following steps:
\begin{enumerate}
\item \underline{Rotation of coordinate system} 
Choose the momenta $p_1$ and $p_2$ of the incoming and outgoing
electron in the frame in which the longitudinal axis is taken along
the photon:

\begin{eqnarray}
  {\bf p}_{1} &=& {-1\over x}\, {\bf k}  \, ,\qquad 
  {\bf p}_{2} = {\bf p} - {1-x\over x}\, {\bf k}\, ,
  \label{b.3}
\end{eqnarray}
\item \underline{z-dependent phase:}
The z-dependent phases of the Furry wave function (\ref{3.22}) combine
in the radiation amplitude (\ref{b.2}) to an inverse photon formation
length
\begin{equation}
  \bar{q} = p_1 - p_2 - k = \frac{x\, m_e^2}{2\, (1-x)\, E_1}\, .
  \label{b.4}
\end{equation}
\item \underline{Simplifying the spinor structure:} 
       The spinor structure 
\begin{equation}
 \widehat\Gamma_r =
 \sqrt{1-x}\,u^*({\bf p}_2)\,\hat D^*_2\,
 {\bf \alpha}\cdot {\bf \epsilon}\,\hat D_1\,u({\bf p}_1)
 \label{b.5}
\end{equation}
in the amplitude (\ref{b.2}) can be simplified on the cross section
level. The spin- and 
helicity-averaged combination $\widehat\Gamma_r\, \widehat\Gamma_{r'}^*$ 
take the simple form
\begin{equation}
  \widehat\Gamma_r\, \widehat\Gamma_{r'}^* =
  \left[ 4-4x+2x^2\right] \frac{\partial}{\partial{\bf r}} \cdot
                          \frac{\partial}{\partial{\bf r'}}\, 
  + 2m_e^2x^2\, .
  \label{b.6}
\end{equation}
\item
\underline{in-medium average:} The cross section (\ref{b.1})
contains products of the Green's functions (\ref{3.30}). These
are averaged over the distribution of scattering centers in
the medium, see (\ref{3.24}) for the non-abelian case. These 
averages can be written in terms of the
dipole cross section (\ref{2.14}): 
 \begin{eqnarray}
   && \Bigg\langle \exp\left\{ i\int\limits_{z}^{z'}{\it d}\xi\,  
    \left[ U\bigl({\bf r}(\xi),\xi\bigr) - U\bigl({\bf r}'(\xi),\xi\bigr) 
           \right]
    \right\}\Bigg\rangle \nonumber \\
   && \qquad = \exp\left\{ - \frac{1}{2} \int\limits_{z}^{z'}{\it d}\xi\,
       n(\xi)\, \sigma\bigr({\bf r}(\xi)-{\bf r}'(\xi)\bigl)  
                             \right\}\, .
   \label{b.7} 
 \end{eqnarray}
For the products of Green's functions, this leads to
\begin{eqnarray}
&&\langle G(\rho,z';{\bf r},z|p_2)\,\,
                  G^*({\bf r}',z';\rho',z|p_1)\, \rangle
   = \frac{-p}{2\pi\, i\, (z'-z)} \frac{1}{x}
 \nonumber \\
 &&
    \times \exp\left\{ \frac{ip}{2\, (z'-z)} 
      \left[ (1-x) \bigl(\rho - {\bf r}\bigr)^2
             -  \bigl({\bf r}' - \rho'\bigr)^2 \right]
             \right\}
 \nonumber \\
 && \times \int {\cal D}{\bf r}_b
    \exp\left\{ \frac{i\mu}{2}\int_z^{z'}{\dot {\bf r}}_b^2
                - \int_z^{z'} \Sigma\bigl(\xi,{\hat \rho}_f 
                + x{\bf r}_b\bigr)
        \right\}\, ,
 \label{b.8}
\end{eqnarray}
where $\mu = p\, (1-x)\, x$ and 
\begin{eqnarray}
    {\hat\rho}_f(\xi) = {\rho}_2
                             \frac{\xi-z}{z'-z}
                           + 
                             {\rho}_1
                             \frac{z'-\xi}{z'-z}\, .
  \label{b.9}
\end{eqnarray}
\end{enumerate}
After inserting the Furry wave function (\ref{3.22})
into the radiation amplitude (\ref{b.2}). The radiation
probability $\langle|M_{fi}|^2\rangle$ in terms of averages of
Green's functions reads 
\begin{eqnarray}
 \langle|M_{fi}|^2\rangle &=& \frac{2}{4\, p_1^2\, (1-x)^2}\, {\rm Re}\, 
 \int {\it d}{\bf r}_1\, {\it d}{\bf r}\, {\it d}\rho\, d{\bf r}_2\,
      {\it d}{\bf r}'_1\, {\it d}{\bf r}'\, 
      {\it d}\rho'\, {\it d}{\bf r}'_2
      \int dz \int\limits_z^\infty dz'
      \nonumber \\
  &&\quad \times 
       e^{-i{\bf p}_{2}\cdot({\bf r}_2-{\bf r}_2')
        +i{\bf p}_{1}\cdot({\bf r}_1-{\bf r}_1')
        -i\bar{q}(z'-z)} e^{-\epsilon (|z|+|z'|)} \nonumber \\
  &&\quad \times 
                  \langle G({\bf r}_2,z_+;\rho,z'|p_2)\,
                  G^*({\bf r}'_2,z_+;{\bf r}',z'|p_2)\rangle
   \nonumber \\
  &&\quad \times \widehat\Gamma_{-r}\, \widehat\Gamma_{r'}^*
      \langle G(\rho,z';{\bf r},z|p_2)\,
                  G^*({\bf r}',z';\rho',z|p_1)\rangle
   \nonumber \\
  &&\quad \times \langle G({\bf r},z;{\bf r}_1,z_-|p_1)\,
                  G^*({\rho}',z;{\bf r}'_1,z_-|p_1)\rangle\, .
  \label{b.10}
\end{eqnarray}
Equation (\ref{b.10}) is rearranged such that the photon emission
in the amplitude ${\cal M}_{\rm fi}$ occurs prior to emission in the
complex conjugate amplitude ${\cal M}_{\rm fi}^*$. The opposite 
contribution is accounted for by taking twice the real part. 
Averages $\langle \dots \rangle$ of pairs of Green's functions 
effectively compare the paths ${\bf r}(\xi)$ and ${\bf r}'(\xi)$ 
of the electron in amplitude and complex conjugate amplitude,
see eq. (\ref{b.8}). Using this average, one arrives at the
final expression\cite{Zakharov:1998uh,Wiedemann:1999fq}
\begin{eqnarray}
 &&\frac{d^3\sigma}{d({\rm ln}x)\,
   d{\bf k}} = \frac{\alpha_{em}}
  {(2\pi)^2}\, {2\over {E_1^2\, (1-x)^2}}\,
     \nonumber \\
 && \qquad \times {\rm Re}\, 
    \int\limits_{z_-}^{z_+}\, dz \int\limits_{z}^{z_+}\, dz' 
        \exp\left\{-i\frac{x\, m_e^2}{2\, (1-x)\, E_1}(z'-z) 
                   -\epsilon(|z|+|z'|)\right\}
    \nonumber \\
  && \qquad \times  \int 
     {\it d}{\bf r}_1
     \exp\left\{ -i{\bf k}\cdot {\bf r}_1 
         \right\} \, 
    \exp\left\{
    -\int\limits_{z_-}^z d\xi\, n(\xi)\, \sigma(x\, {\bf r}_1) 
    \right\}
     \nonumber \\ 
 && \qquad 
    \times 
    \left[ {{4-4x+2x^2}\over 4x^2} {\partial\over \partial{\bf r}_1}
                           \cdot {\partial\over \partial{\bf r}_2}
        + {m_e^2\, x^2\over 2} \right] 
        {\cal K}\bigl(z',{\bf r}_2=0;z,{\bf r}_1|\mu\bigr) \, ,
    \label{b.11} 
\end{eqnarray}
where the abelian path-integral takes the form
\begin{eqnarray}
  &&{\cal K}\bigl(z',{\bf r}_c(z');z,{\bf r}_c(z)|\mu = E_1(1-x)x\bigr) 
  \nonumber \\
  && \quad =   \int {\cal D}{\bf r}_c\, 
  \exp\left\{i\, \int\limits_{z}^{z'}\, d\xi\,
  \left[{\mu\over 2}\dot{\bf r}_c^2 
  + i\, \Sigma\bigl(\xi,x\, {\bf r}_c\bigr)\right] \right\}\, .
  \label{b.12}
\end{eqnarray}
The expression (\ref{b.11}) provides a path-integral
formulation of the Landau-Pomeranchuk-Migdal radiation 
spectrum\cite{Landau:um,Landau:gr,Migdal:1956tc,Klein:1998du}. 
Graphically, as can be seen from eq. (\ref{b.10}), this 
radiation spectrum can be represented as\\

\begin{figure}[h]\epsfxsize=8.5cm 
\centerline{\epsfbox{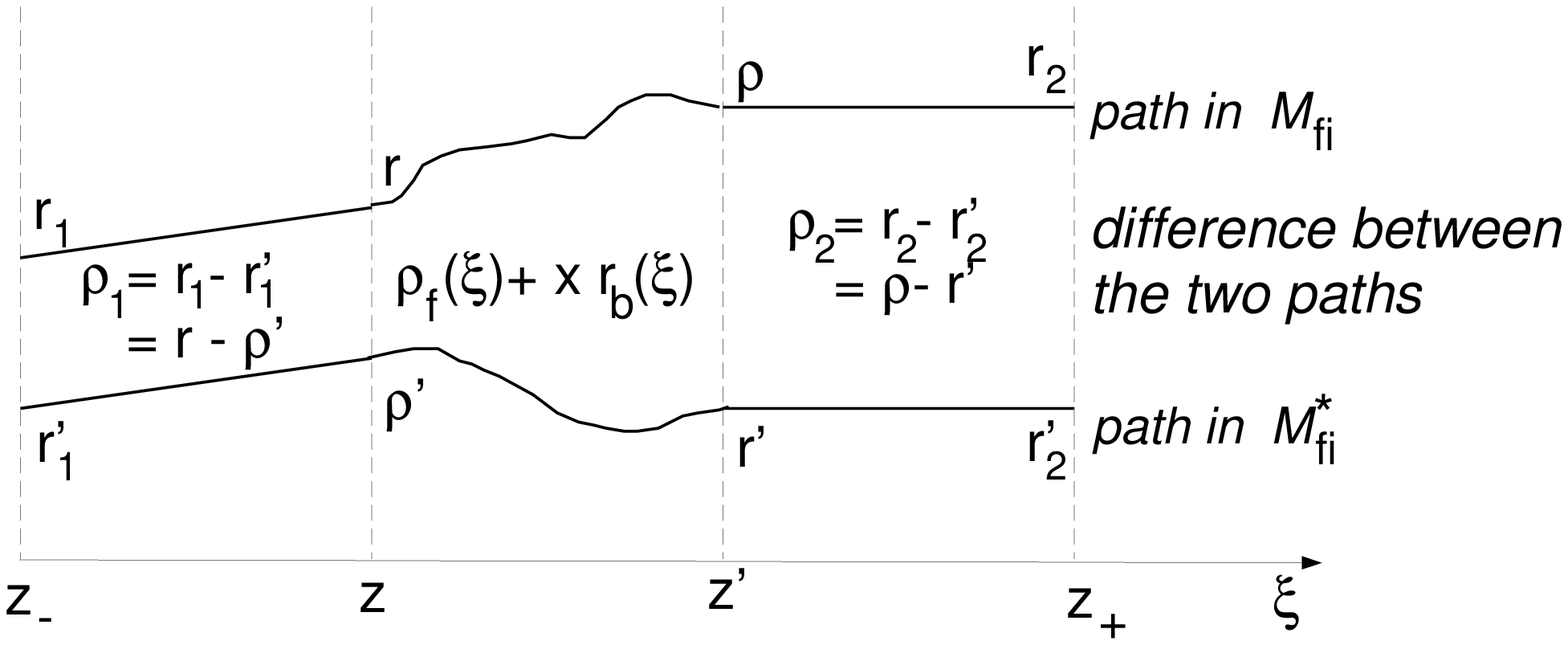}}
\end{figure}
%
\noindent
Thus, for propagation from $z_- = -\infty$ up to the longitudinal
photon emission point $z$ in ${\cal M}_{\rm fi}$,
the electron propagates with initial energy $E_1$ in both
amplitude and complex amplitude. Then it propagates with 
momentum $E_2 = E_1(1-x)$ in ${\cal M}_{\rm fi}$ but with $E_1$ in 
${\cal M}_{\rm fi}^*$ up to $z'$, the photon emission point
in the complex amplitude, etc. The average (\ref{b.8}) ensures
that if propagation in ${\cal M}_{\rm fi}$ and ${\cal M}_{\rm fi}^*$
occur with the same energy, then the relative distance between 
the paths ${\bf r}(\xi)$ and ${\bf r}'(\xi)$ does not change:
physically, the radiation cross section counts only those
changes in ${\bf r}(\xi)-{\bf r}'(\xi)$ which amount to 
phase shifts between amplitude and complex conjugate amplitude.
Those are accumulated between the photon emission points $z$ and
$z'$ in ${\cal M}_{\rm fi}$ and ${\cal M}_{\rm fi}^*$, respectively.

\section*{Acknowledgments}
Both authors thank the Institute for Nuclear Theory at the University
of Washington for its hospitality and the Department of Energy for
partial support during the completion of this work.


\end{document}